\documentclass[aps,preprint]{revtex4}%
\usepackage{amsfonts}
\usepackage{amsmath}
\usepackage{amssymb}
\usepackage{graphicx}%
\begin{document}
\preprint{ }

\vspace*{1cm}

\begin{center}

{\Large Chiral effective field theory on the lattice at next-to-leading
order}\vspace*{0.75cm}

{Bu\={g}ra~Borasoy$^{a}$, Evgeny Epelbaum$^{b,a}$, Hermann~Krebs$^{a,b}$,
Dean~Lee$^{c,a}$, Ulf-G.~Mei{\ss }ner$^{a,b}$}\vspace*{0.75cm}

$^{a}$\textit{Helmholtz-Institut f\"{u}r Strahlen- und Kernphysik (Theorie)
Universit\"{a}t Bonn, }\linebreak\textit{Nu\ss allee 14-16, D-53115 Bonn,
Germany }

$^{b}$\textit{Institut f\"{u}r Kernphysik (Theorie), Forschungszentrum
J\"{u}lich, D-52425 J\"{u}lich, Germany }

$^{c}$\textit{Department of Physics, North Carolina State University, Raleigh,
NC 27695, USA}

\vspace*{0.75cm}

{\large Abstract}
\end{center}

We study nucleon-nucleon scattering on the lattice at next-to-leading order in
chiral effective field theory. \ We determine phase shifts and mixing angles
from the properties of two-nucleon standing waves induced by a hard spherical
wall in the center-of-mass frame. \ At fixed lattice spacing we test model
independence of the low-energy effective theory by computing
next-to-leading-order corrections for two different leading-order lattice
actions. \ The first leading-order action includes instantaneous one-pion
exchange and same-site contact interactions. \ The second leading-order action
includes instantaneous one-pion exchange and Gaussian-smeared interactions.
\ We find that in each case the results at next-to-leading order are accurate
up to corrections expected at higher order.\pagebreak

\section{Introduction}

There have been several recent studies on the subject of lattice simulations
for low-energy nuclear physics using effective interactions
\cite{Muller:1999cp,Abe:2003fz,Chandrasekharan:2003ub,Chandrasekharan:2003wy,Lee:2004si,Lee:2004qd,Hamilton:2004gp,Seki:2005ns,Lee:2005is,Lee:2005it,Borasoy:2005yc,deSoto:2006pe,Borasoy:2006qn,Lee:2007jd,Abe:2007fe,Abe:2007ff}%
. \ \ \ Here we present a study of nucleon-nucleon scattering on the lattice
at next-to-leading order in chiral effective field theory for momenta less
than or equal to the pion mass. \ This analysis is part of a long-term effort
to put the formalism of chiral effective field theory on the lattice and use
the lattice action to study few- and many-nucleon systems. \ This is the first
of a pair of papers on lattice chiral effective field theory at
next-to-leading order. \ In this first paper we calculate lattice phase shifts
and mixing angles and fit unknown operator coefficients at next-to-leading
order using the spherical wall method introduced in \cite{Borasoy:2007vy}.
\ In the companion paper we use the resulting lattice action to perform Monte
Carlo simulations of dilute neutron matter.

The organization of this paper is as follows. \ We begin with a review of the
effective potential for chiral effective field theory at next-to-leading order
and simplifications which can be made at low cutoff momentum. \ We then
discuss a general test of model independence at fixed lattice spacing. \ We
illustrate this model independence explicitly by repeating all calculations
using two different lattice actions. \ The first action uses instantaneous
one-pion exchange and same-site contact interactions at leading order. \ The
second leading-order action includes instantaneous one-pion exchange and
Gaussian-smeared interactions. \ Both of these lattice actions were first
introduced in \cite{Borasoy:2006qn}. \ We compute next-to-leading-order
corrections for these actions and find results accurate up to omitted
higher-order terms.

\section{Chiral effective field theory at next-to-leading order}

\subsection{Effective potential}

In the following $\vec{q}$ denotes the $t$-channel momentum transfer for
nucleon-nucleon scattering while $\vec{k}$ is the $u$-channel exchanged
momentum transfer. At leading order (LO) in the Weinberg power-counting scheme
\cite{Weinberg:1990rz,Weinberg:1991um} the $NN$ effective potential consists
of two independent contact terms and instantaneous one-pion exchange (OPEP),%
\begin{equation}
V_{\text{LO}}=V^{(0)}+V^{\text{OPEP}}, \label{VLO}%
\end{equation}%
\begin{equation}
V^{(0)}=C_{S}+C_{T}\left(  \vec{\sigma}_{1}\cdot\vec{\sigma}_{2}\right)  ,
\label{V0}%
\end{equation}%
\begin{equation}
V^{\text{OPEP}}=-\left(  \frac{g_{A}}{2f_{\pi}}\right)  ^{2}\boldsymbol\tau
_{1}\cdot\boldsymbol\tau_{2}\frac{\left(  \vec{\sigma}_{1}\cdot\vec{q}\right)
\left(  \vec{\sigma}_{2}\cdot\vec{q}\right)  }{q^{\,2}+m_{\pi}^{2}}.
\label{VOPEP}%
\end{equation}
The vector arrow in $\vec{\sigma}$ signifies the three-vector index for spin.
\ The boldface for $\boldsymbol\tau$ signifies the three-vector index for
isospin. \ We take for our physical constants $m=938.92$ MeV as the nucleon
mass, $m_{\pi}=138.08$ MeV as the pion mass, $f_{\pi}=93$ MeV as the pion
decay constant, and $g_{A}=1.26$ as the nucleon axial charge.

At next-to-leading order (NLO) the effective potential introduces seven
independent contact terms carrying two powers of momentum, corrections to the
two LO\ contact terms, and instantaneous two-pion exchange (TPEP)
\cite{Ordonez:1992xp,Ordonez:1993tn,Ordonez:1996rz,Epelbaum:1998ka,Epelbaum:1999dj}%
. \ Using the notational conventions of \cite{Epelbaum:1998ka,Epelbaum:1999dj}
we have%
\begin{equation}
V_{\text{NLO}}=V_{\text{LO}}+\Delta V^{(0)}+V^{(2)}+V_{\text{NLO}%
}^{\text{TPEP}}. \label{VNLO}%
\end{equation}
The contact interactions are given by%
\begin{equation}
\Delta V^{(0)}=\Delta C_{S}+\Delta C_{T}\left(  \vec{\sigma}_{1}\cdot
\vec{\sigma}_{2}\right)  , \label{dV0}%
\end{equation}%
\begin{align}
V^{(2)}  &  =C_{1}q^{2}+C_{2}k^{2}+\left(  C_{3}q^{2}+C_{4}k^{2}\right)
\left(  \vec{\sigma}_{1}\cdot\vec{\sigma}_{2}\right)  +iC_{5}\frac{1}%
{2}\left(  \vec{\sigma}_{1}+\vec{\sigma}_{2}\right)  \cdot\left(  \vec
{q}\times\vec{k}\right) \nonumber\\
&  +C_{6}\left(  \vec{\sigma}_{1}\cdot\vec{q}\right)  \left(  \vec{\sigma}%
_{2}\cdot\vec{q}\right)  +C_{7}\left(  \vec{\sigma}_{1}\cdot\vec{k}\right)
\left(  \vec{\sigma}_{2}\cdot\vec{k}\right)  , \label{V2}%
\end{align}
and the NLO two-pion exchange potential is \cite{Friar:1994,Kaiser:1997mw}%
\begin{align}
V_{\text{NLO}}^{\text{TPEP}}  &  =-\frac{\boldsymbol\tau_{1}\cdot
\boldsymbol\tau_{2}}{384\pi^{2}f_{\pi}^{4}}L(q)\left[  4m_{\pi}^{2}\left(
5g_{A}^{4}-4g_{A}^{2}-1\right)  +q^{2}\left(  23g_{A}^{4}-10g_{A}%
^{2}-1\right)  +\frac{48g_{A}^{4}m_{\pi}^{4}}{4m_{\pi}^{2}+q^{2}}\right]
\nonumber\\
&  -\frac{3g_{A}^{4}}{64\pi^{2}f_{\pi}^{4}}L(q)\left[  \left(  \vec{q}%
\cdot\vec{\sigma}_{1}\right)  \left(  \vec{q}\cdot\vec{\sigma}_{2}\right)
-q^{2}\left(  \vec{\sigma}_{1}\cdot\vec{\sigma}_{2}\right)  \right]  ,
\label{VTPEPNLO}%
\end{align}
where%
\begin{equation}
L(q)=\frac{1}{2q}\sqrt{4m_{\pi}^{2}+q^{2}}\ln\frac{\sqrt{4m_{\pi}^{2}+q^{2}%
}+q}{\sqrt{4m_{\pi}^{2}+q^{2}}-q}. \label{Lq}%
\end{equation}

\subsection{Simplified form at low cutoff}

In this paper we consider low-energy nucleon-nucleon scattering at momenta
less than or equal to the pion mass, $m_{\pi}$. \ On the lattice the
ultaviolet cutoff momentum, $\Lambda$, equals $\pi$ divided by the lattice
spacing, $a$. \ If we focus narrowly on calculations of two-nucleon scattering
on the lattice we can take any sufficiently small lattice spacing satisfying
$\Lambda\gg m_{\pi}$.\ \ However serious numerical difficulties appear at
large $\Lambda$ in Monte Carlo simulations of few- and many-nucleon systems.
\ In some attractive channels we might find spurious deeply-bound states at
large $\Lambda$. \ In other channels we must deal with short-range hard-core
repulsion becoming more prominent and producing fluctuating sign or complex
phase cancellations. \ The general connection between sign/phase oscillations
and repulsive interactions has been discussed in the literature in several
different contexts \cite{Loh:1990,Koonin:1997,Chen:2004rq}. \ The severity of
the sign/phase problem scales exponentially with system size and strength of
the repulsive interaction.

In order to avoid these difficulties we set the cutoff momentum $\Lambda$ as
low as possible for describing physical momenta up to $m_{\pi}$. \ We take
$\Lambda=314$ MeV $\approx2.3m_{\pi}$, corresponding with $a^{-1}=100$ MeV.
\ \ This coarse lattice approach is similar in motivation to the continuum
low-momentum potential $V_{\text{low }k}$ derived using the renormalization
group \cite{Bogner:2001gq,Bogner:2003wn}. \ There has been some discussion in
the literature about the consistency of the Weinberg power counting scheme at
high momentum cutoff, starting with the work of
\cite{Kaplan:1996xu,Kaplan:1998we} and more recently \cite{Beane:2001bc,
Nogga:2005hy, Birse:2005um, Epelbaum:2006pt, Birse:2007sx}. \ Due to the
computational reasons explained above, we are prevented from reaching high
momentum cutoff scales in our lattice calculations where alternative power
counting schemes may be useful.

For nearly all $\left\vert q\right\vert <\Lambda$ we can expand the two-pion
exchange potential in powers of $q^{2}/(4m_{\pi}^{2}),$%
\begin{equation}
L(q)=1+\frac{1}{3}\frac{q^{2}}{4m_{\pi}^{2}}+\cdots,
\end{equation}%
\begin{equation}
\frac{4m_{\pi}^{2}}{4m_{\pi}^{2}+q^{2}}L(q)=1-\frac{2}{3}\frac{q^{2}}{4m_{\pi
}^{2}}+\cdots,
\end{equation}%
\begin{align}
V_{\text{NLO}}^{\text{TPEP}}  &  =-\frac{\boldsymbol\tau_{1}\cdot
\boldsymbol\tau_{2}}{384\pi^{2}f_{\pi}^{4}}\left[  4m_{\pi}^{2}\left(
8g_{A}^{4}-4g_{A}^{2}-1\right)  +\frac{2}{3}q^{2}\left(  34g_{A}^{4}%
-17g_{A}^{2}-2\right)  +O\left(  \left(  \tfrac{q^{2}}{4m_{\pi}^{2}}\right)
^{2}\right)  \right] \nonumber\\
&  -\frac{3g_{A}^{4}}{64\pi^{2}f_{\pi}^{4}}\left[  \left(  \vec{q}\cdot
\vec{\sigma}_{1}\right)  \left(  \vec{q}\cdot\vec{\sigma}_{2}\right)
-q^{2}\left(  \vec{\sigma}_{1}\cdot\vec{\sigma}_{2}\right)  \right]  \left[
1+O\left(  \tfrac{q^{2}}{4m_{\pi}^{2}}\right)  \right]  . \label{localTPEP}%
\end{align}
This expansion fails to converge only for values of $q$ near the cutoff scale
$\Lambda$ $\approx2.3m_{\pi}$, where the effective theory is already
problematic due to large cutoff effects of size $O\left(  q^{2}/\Lambda
^{2}\right)  $. \ From a practical viewpoint there is no need to keep the full
non-local structure of $V_{\text{NLO}}^{\text{TPEP}}$ at this lattice spacing.
\ Instead we simply use%
\begin{equation}
V_{\text{LO}}=V^{(0)}+V^{\text{OPEP}},
\end{equation}%
\begin{equation}
V_{\text{NLO}}=V_{\text{LO}}+\Delta V^{(0)}+V^{(2)},
\end{equation}
where the terms in Eq.~(\ref{localTPEP}) with up to two powers of $q$ are
absorbed as a redefinition of the coefficients $\Delta V^{(0)}$ and $V^{(2)}$.
\ This same approach can be applied to the two-pion exchange potential at
next-to-next-to-leading order (NNLO) and higher-order $n$-pion exchange potentials.

\subsection{Model independence at fixed lattice spacing}

The usual test of model independence in low-energy effective field theory
calculations is to check for sensitivity on the cutoff scale $\Lambda$. \ The
difference between calculations at a given order for two different cutoff
scales $\Lambda_{1}$ and $\Lambda_{2}$ should be no larger than the omitted
corrections at the next order. \ On the lattice this test is problematic since
we cannot change the lattice spacing by a large amount due to computational
constraints. \ Fortunately cutoff independence is just one of many ways to
test model independence. In the following we discuss a general class of
methods to check model independence at fixed lattice spacing.

Let us use the notation $V^{Q^{n}/\Lambda^{n}}$ to denote two-nucleon
operators with the following properties. \ $V^{Q^{n}/\Lambda^{n}}$ is a sum of
local two-nucleon interactions each with at least $n$ powers of momenta and an
analytic function of momenta below the cutoff scale $\Lambda$. \ We use the
adjective \textquotedblleft quasi-local\textquotedblright\ to describe
$V^{Q^{n}/\Lambda^{n}}$ since the interactions are purely short range.
\ Quasi-local operators of this type arise naturally in improved lattice
actions of the type considered in \cite{Borasoy:2006qn}.

At fixed lattice spacing we consider two modified lowest-order actions of the
form%
\begin{align}
V_{\text{LO}_{1}}  &  =V_{1}^{(0)}+V^{\text{OPEP}}+V_{1}^{Q^{2}/\Lambda^{2}%
},\label{LO1}\\
V_{\text{LO}_{2}}  &  =V_{2}^{(0)}+V^{\text{OPEP}}+V_{2}^{Q^{2}/\Lambda^{2}},
\label{LO2}%
\end{align}
where $V_{1}^{Q^{2}/\Lambda^{2}}$ and $V_{2}^{Q^{2}/\Lambda^{2}}$ are
different quasi-local operators with at least two powers of momenta. \ The
leading-order interactions are iterated nonperturbatively and so, when fitted
to physical low-energy data, the contact terms $V_{1}^{(0)}$ and $V_{2}^{(0)}$
in general have different coefficients. \ However we expect low-energy
physical observables such as scattering phase shifts, binding energies, etc.,
should agree up to differences comparable to the omitted NLO corrections.

Similarly at NLO we may consider modified actions of the form%
\begin{align}
V_{\text{NLO}_{1}} &  =V_{\text{LO}_{1}}+\Delta V_{1}^{(0)}+V_{1}^{(2)}%
+V_{1}^{Q^{4}/\Lambda^{4}},\\
V_{\text{NLO}_{2}} &  =V_{\text{LO}_{2}}+\Delta V_{2}^{(0)}+V_{2}^{(2)}%
+V_{2}^{Q^{4}/\Lambda^{4}},
\end{align}
where $V_{1}^{Q^{4}/\Lambda^{4}}$ and $V_{2}^{Q^{4}/\Lambda^{4}}$ are
different quasi-local operators with at least four powers of momenta. \ Once
again low-energy physical observables should agree up to differences
comparable to the omitted corrections at the next order.

At any given order the effect of small changes of the lattice spacing can be
reinterpreted at fixed lattice spacing as a renormalization group
transformation on higher-order local operator coefficients. \ In principle
this type of modification is covered by our condition of model independence
for general $V^{Q^{n}/\Lambda^{n}}$. \ Nevertheless it is good to check
explicitly as many variations as possible, and studies at different lattice
spacings as well as different functional forms for $V^{Q^{n}/\Lambda^{n}}$ are
planned for the future.

\section{Lattice formalism}

\subsection{Lattice notation and cubic symmetries}

In this paper we assume exact isospin symmetry and neglect electromagnetic
interactions. \ We use $\vec{n}$ to represent integer-valued lattice vectors
on a three-dimensional spatial lattice and either $\vec{p},$ $\vec{q}$, or
$\vec{k}$ to represent integer-valued momentum lattice vectors.$\ \ \hat
{l}=\hat{1}$, $\hat{2}$, $\hat{3}$ are unit lattice vectors in the spatial
directions, $a$ is the spatial lattice spacing, and $L$ is the length of the
cubic spatial lattice in each direction. \ We use the Euclidean transfer
matrix formalism defined in \cite{Borasoy:2006qn} with lattice time step
$a_{t}$, and the integer $n_{t}$ labels the time steps. \ We define
$\alpha_{t}$ as the ratio between lattice spacings, $\alpha_{t}=a_{t}/a$.
\ Throughout we use dimensionless parameters and operators, which correspond
with physical values multiplied by the appropriate power of $a$. \ Final
results are presented in physical units with the corresponding unit stated
explicitly. \ As in \cite{Borasoy:2006qn} we use the spatial lattice spacing
$a=(100$ MeV$)^{-1}$ and temporal lattice spacing $a_{t}=(70$ MeV$)^{-1}$.

We use $a$ and $a^{\dagger}$ to denote annihilation and creation operators.
\ To avoid confusion we make explicit in our lattice notation all spin and
isospin indices using%
\begin{align}
a_{0,0}  &  =a_{\uparrow,p},\text{ \ }a_{0,1}=a_{\uparrow,n},\\
a_{1,0}  &  =a_{\downarrow,p},\text{ \ }a_{1,1}=a_{\downarrow,n}.
\end{align}
The first subscript is for spin and the second subscript is for isospin. \ We
use $\tau_{I}$ with $I=1,2,3$ to represent Pauli matrices acting in isospin
space and $\sigma_{S}$ with $S=1,2,3$ to represent Pauli matrices acting in
spin space. \ We also use the letters $S$ and $I$ to denote the total spin and
total isospin for the two-nucleon system. \ The intended meaning in each case
should be clear from the context.

On the lattice the rotational symmetry is reduced to the cubic subgroup
SO$(3,\mathbb{Z})$ of SO$(3)$ while isospin symmetry remains intact as the
full SU(2) symmetry. \ There are five irreducible representations of the cubic
rotational group. \ These are usually written as $A_{1}$, $T_{1}$, $E$,
$T_{2}$, and $A_{2}$. \ Some of their properties and examples in terms of
spherical harmonics $Y_{L,L_{z}}(\theta,\phi)$ are listed in Table \ref{reps}.
\ \begin{table}[tb]
\caption{Irreducible SO$(3,\mathbb{Z})$ representations}
$%
\begin{tabular}
[c]{||c|c|c||}\hline\hline
Re$\text{presentation}$ & $J_{z}$ & Ex$\text{ample}$\\\hline
$A_{1}$ & $0\operatorname{mod}4$ & $Y_{0,0}$\\\hline
$T_{1}$ & $0,1,3\operatorname{mod}4$ & $\left\{  Y_{1,0},Y_{1,1},Y_{1,-1}
\right\}  $\\\hline
$E$ & $0,2\operatorname{mod}4$ & $\left\{  Y_{2,0},\frac{Y_{2,-2}+Y_{2,2}%
}{\sqrt{2}}\right\}  $\\\hline
$T_{2}$ & $1,2,3\operatorname{mod}4$ & $\left\{  Y_{2,1},\frac{Y_{2,-2}%
-Y_{2,2}}{\sqrt{2}},Y_{2,-1}\right\}  $\\\hline
$A_{2}$ & $2\operatorname{mod}4$ & $\frac{Y_{3,2}-Y_{3,-2}}{\sqrt{2}}%
$\\\hline\hline
\end{tabular}
$\label{reps}%
\end{table}The $2J+1$ elements of the total angular momentum $J$
representation of SO$(3)$ break up into smaller pieces consisting of the five
irreducible representations. \ 

We use the eight vertices of a unit cube on the lattice to define spatial
derivatives. \ For each spatial direction $l=1,2,3$ and any lattice function
$f(\vec{n})$, let%
\begin{equation}
\Delta_{l}f(\vec{n})=\frac{1}{4}\sum_{\substack{\nu_{1},\nu_{2},\nu_{3}%
=0,1}}(-1)^{\nu_{l}+1}f(\vec{n}+\vec{\nu}),\qquad\vec{\nu}=\nu_{1}\hat{1}%
+\nu_{2}\hat{2}+\nu_{3}\hat{3}. \label{derivative}%
\end{equation}
We also define the double spatial derivative along direction $l$,%
\begin{equation}
\triangledown_{l}^{2}f(\vec{n})=f(\vec{n}+\hat{l})+f(\vec{n}-\hat{l}%
)-2f(\vec{n}).
\end{equation}

\subsection{Densities and current densities}

We define the local density,%
\begin{equation}
\rho^{a^{\dagger},a}(\vec{n})=\sum_{i,j=0,1}a_{i,j}^{\dagger}(\vec{n}%
)a_{i,j}(\vec{n}),
\end{equation}
which is invariant under Wigner's SU(4) symmetry \cite{Wigner:1937}.
\ Similarly we define the local spin density for $S=1,2,3,$%
\begin{equation}
\rho_{S}^{a^{\dagger},a}(\vec{n})=\sum_{i,j,i^{\prime}=0,1}a_{i,j}^{\dagger
}(\vec{n})\left[  \sigma_{S}\right]  _{ii^{\prime}}a_{i^{\prime},j}(\vec{n}),
\end{equation}
isospin density for $I=1,2,3,$%
\begin{equation}
\rho_{I}^{a^{\dagger},a}(\vec{n})=\sum_{i,j,j^{\prime}=0,1}a_{i,j}^{\dagger
}(\vec{n})\left[  \tau_{I}\right]  _{jj^{\prime}}a_{i,j^{\prime}}(\vec{n}),
\end{equation}
and spin-isospin density for $S,I=1,2,3,$%
\begin{equation}
\rho_{S,I}^{a^{\dagger},a}(\vec{n})=\sum_{i,j,i^{\prime},j^{\prime}%
=0,1}a_{i,j}^{\dagger}(\vec{n})\left[  \sigma_{S}\right]  _{ii^{\prime}%
}\left[  \tau_{I}\right]  _{jj^{\prime}}a_{i^{\prime},j^{\prime}}(\vec{n}).
\end{equation}

For each static density we also have an associated current density. \ Similar
to the definition of the lattice derivative $\Delta_{l}$ in
Eq.~(\ref{derivative}), we use the eight vertices of a unit cube,
\begin{equation}
\vec{\nu}=\nu_{1}\hat{1}+\nu_{2}\hat{2}+\nu_{3}\hat{3},
\end{equation}
for $\nu_{1},\nu_{2},\nu_{3}=0,1$. \ Let $\vec{\nu}(-l)$ for $l=1,2,3$ be the
result of reflecting the $l^{\text{th}}$-component of $\vec{\nu}$ about the
center of the cube,%
\begin{equation}
\vec{\nu}(-l)=\vec{\nu}+(1-2\nu_{l})\hat{l}.
\end{equation}
Omitting factors of $i$ and $1/m$, we can write the $l^{\text{th}}$-component
of the SU(4)-invariant current density as%
\begin{equation}
\Pi_{l}^{a^{\dagger},a}(\vec{n})=\frac{1}{4}\sum_{\substack{\nu_{1},\nu
_{2},\nu_{3}=0,1}}\sum_{i,j=0,1}(-1)^{\nu_{l}+1}a_{i,j}^{\dagger}(\vec{n}%
+\vec{\nu}(-l))a_{i,j}(\vec{n}+\vec{\nu}).
\end{equation}
Similarly the $l^{\text{th}}$-component of spin current density is%
\begin{equation}
\Pi_{l,S}^{a^{\dagger},a}(\vec{n})=\frac{1}{4}\sum_{\substack{\nu_{1},\nu
_{2},\nu_{3}=0,1}}\sum_{i,j,i^{\prime}=0,1}(-1)^{\nu_{l}+1}a_{i,j}^{\dagger
}(\vec{n}+\vec{\nu}(-l))\left[  \sigma_{S}\right]  _{ii^{\prime}}a_{i^{\prime
},j}(\vec{n}+\vec{\nu}),
\end{equation}
$l^{\text{th}}$-component of isospin current density is%
\begin{equation}
\Pi_{l,I}^{a^{\dagger},a}(\vec{n})=\frac{1}{4}\sum_{\substack{\nu_{1},\nu
_{2},\nu_{3}=0,1}}\sum_{i,j,j^{\prime}=0,1}(-1)^{\nu_{l}+1}a_{i,j}^{\dagger
}(\vec{n}+\vec{\nu}(-l))\left[  \tau_{I}\right]  _{jj^{\prime}}a_{i,j^{\prime
}}(\vec{n}+\vec{\nu}),
\end{equation}
and $l^{\text{th}}$-component of spin-isospin current density is%
\begin{equation}
\Pi_{l,S,I}^{a^{\dagger},a}(\vec{n})=\frac{1}{4}\sum_{\substack{\nu_{1}%
,\nu_{2},\nu_{3}=0,1}}\sum_{i,j,i^{\prime},j^{\prime}=0,1}(-1)^{\nu_{l}%
+1}a_{i,j}^{\dagger}(\vec{n}+\vec{\nu}(-l))\left[  \sigma_{S}\right]
_{ii^{\prime}}\left[  \tau_{I}\right]  _{jj^{\prime}}a_{i^{\prime},j^{\prime}%
}(\vec{n}+\vec{\nu}).
\end{equation}

\subsection{Instantaneous free pion action}

The lattice action for free pions with purely instantaneous propagation is%
\begin{equation}
S_{\pi\pi}(\pi_{I})=\alpha_{t}(\tfrac{m_{\pi}^{2}}{2}+3)\sum_{\vec{n},n_{t}%
,I}\pi_{I}(\vec{n},n_{t})\pi_{I}(\vec{n},n_{t})-\alpha_{t}\sum_{\vec{n}%
,n_{t},I,l}\pi_{I}(\vec{n},n_{t})\pi_{I}(\vec{n}+\hat{l},n_{t}),
\end{equation}
where $\pi_{I}$ is the pion field labelled with isospin index $I$. \ We note
that pion fields at different time steps $n_{t}$ and $n_{t}^{\prime}$ are
decoupled due to the omission of time derivatives. \ This generates
instantaneous propagation\ at each time step when computing one-pion exchange
diagrams. \ It is convenient to define a rescaled pion field, $\pi_{I}%
^{\prime}$,%
\begin{equation}
\pi_{I}^{\prime}(\vec{n},n_{t})=\sqrt{q_{\pi}}\pi_{I}(\vec{n},n_{t}),
\end{equation}%
\begin{equation}
q_{\pi}=\alpha_{t}(m_{\pi}^{2}+6).
\end{equation}
Then%
\begin{equation}
S_{\pi\pi}(\pi_{I}^{\prime})=\frac{1}{2}\sum_{\vec{n},n_{t},I}\pi_{I}^{\prime
}(\vec{n},n_{t})\pi_{I}^{\prime}(\vec{n},n_{t})-\frac{\alpha_{t}}{q_{\pi}}%
\sum_{\vec{n},n_{t},I,l}\pi_{I}^{\prime}(\vec{n},n_{t})\pi_{I}^{\prime}%
(\vec{n}+\hat{l},n_{t}).
\end{equation}

In momentum space the action is%
\begin{equation}
S_{\pi\pi}(\pi_{I}^{\prime})=\frac{1}{L^{3}}\sum_{I,\vec{k}}\pi_{I}^{\prime
}(-\vec{k},n_{t})\pi_{I}^{\prime}(\vec{k},n_{t})\left[  \frac{1}{2}%
-\frac{\alpha_{t}}{q_{\pi}}\sum_{l}\cos\left(  \tfrac{2\pi k_{l}}{L}\right)
\right]  .
\end{equation}
The instantaneous pion correlation function at spatial separation $\vec{n}$ is%
\begin{align}
\left\langle \pi_{I}^{\prime}(\vec{n},n_{t})\pi_{I}^{\prime}(\vec{0}%
,n_{t})\right\rangle  &  =\frac{\int D\pi_{I}^{\prime}\;\pi_{I}^{\prime}%
(\vec{n},n_{t})\pi_{I}^{\prime}(\vec{0},n_{t})\;\exp\left[  -S_{\pi\pi
}\right]  }{\int D\pi_{I}^{\prime}\;\exp\left[  -S_{\pi\pi}\right]  }\text{
\ (no sum on }I\text{)}\nonumber\\
&  =\frac{1}{L^{3}}\sum_{\vec{k}}e^{-i\frac{2\pi}{L}\vec{k}\cdot\vec{n}}%
D_{\pi}(\vec{k}),
\end{align}
where%
\begin{equation}
D_{\pi}(\vec{k})=\frac{1}{1-\tfrac{2\alpha_{t}}{q_{\pi}}\sum_{l}\cos\left(
\tfrac{2\pi k_{l}}{L}\right)  }.
\end{equation}
It is useful also to define the two-derivative pion correlator, $G_{S_{1}%
S_{2}}(\vec{n})$,%
\begin{align}
G_{S_{1}S_{2}}(\vec{n})  &  =\left\langle \Delta_{S_{1}}\pi_{I}^{\prime}%
(\vec{n},n_{t})\Delta_{S_{2}}\pi_{I}^{\prime}(\vec{0},n_{t})\right\rangle
\text{ \ (no sum on }I\text{)}\nonumber\\
&  =\frac{1}{16}\sum_{\nu_{1},\nu_{2},\nu_{3}=0,1}\sum_{\nu_{1}^{\prime}%
,\nu_{2}^{\prime},\nu_{3}^{\prime}=0,1}(-1)^{\nu_{S_{1}}}(-1)^{\nu_{S_{2}%
}^{\prime}}\left\langle \pi_{I}^{\prime}(\vec{n}+\vec{\nu}-\vec{\nu}^{\prime
},n_{t})\pi_{I}^{\prime}(\vec{0},n_{t})\right\rangle .
\end{align}

\section{Lattice transfer matrices}

\subsection{Leading-order actions LO$_{1}$ and LO$_{2}$}

The analysis in \cite{Borasoy:2006qn} considers the lattice path integral with
and without auxiliary fields as well as the lattice transfer matrix with and
without auxiliary fields. \ All four formulations are shown to be exactly the
same. Here we discuss only the lattice transfer matrix without auxiliary
fields. \ This formulation is the most useful for calculating nucleon-nucleon
scattering phase shifts and mixing angles. \ In simple terms the Euclidean
transfer matrix is the exponential of the Hamiltonian $\exp(-H\Delta t)$,
where $\Delta t$ equals one temporal lattice spacing. \ For example we can
write the free-nucleon transfer matrix as%
\begin{equation}
M_{\text{free}}\equiv\colon\exp\left(  -H_{\text{free}}\alpha_{t}\right)
\colon,
\end{equation}
where the $::$ symbols indicate normal ordering. \ We use the $O(a^{4}%
)$-improved free lattice Hamiltonian,%
\begin{align}
H_{\text{free}}  &  =\frac{49}{12m}\sum_{\vec{n}}\sum_{i,j=0,1}a_{i,j}%
^{\dagger}(\vec{n})a_{i,j}(\vec{n})\nonumber\\
&  -\frac{3}{4m}\sum_{\vec{n}}\sum_{i,j=0,1}\sum_{l=1,2,3}\left[
a_{i,j}^{\dagger}(\vec{n})a_{i,j}(\vec{n}+\hat{l})+a_{i,j}^{\dagger}(\vec
{n})a_{i,j}(\vec{n}-\hat{l})\right] \nonumber\\
&  +\frac{3}{40m}\sum_{\vec{n}}\sum_{i,j=0,1}\sum_{l=1,2,3}\left[
a_{i,j}^{\dagger}(\vec{n})a_{i,j}(\vec{n}+2\hat{l})+a_{i,j}^{\dagger}(\vec
{n})a_{i,j}(\vec{n}-2\hat{l})\right] \nonumber\\
&  -\frac{1}{180m}\sum_{\vec{n}}\sum_{i,j=0,1}\sum_{l=1,2,3}\left[
a_{i,j}^{\dagger}(\vec{n})a_{i,j}(\vec{n}+3\hat{l})+a_{i,j}^{\dagger}(\vec
{n})a_{i,j}(\vec{n}-3\hat{l})\right]  .
\end{align}

With the interactions included we take the standard lattice transfer matrix
defined in \cite{Borasoy:2006qn} for LO$_{1}$,%
\begin{align}
M_{\text{LO}_{1}}  &  \equiv\colon\exp\left\{  -H_{\text{free}}\alpha
_{t}-\frac{1}{2}C\alpha_{t}\sum_{\vec{n}}\left[  \rho^{a^{\dag},a}(\vec
{n})\right]  ^{2}-\frac{1}{2}C_{I^{2}}\alpha_{t}\sum_{I}\sum_{\vec{n}}\left[
\rho_{I}^{a^{\dag},a}(\vec{n})\right]  ^{2}\right. \nonumber\\
&  +\left.  \frac{g_{A}^{2}\alpha_{t}^{2}}{8f_{\pi}^{2}q_{\pi}}\sum
_{\substack{S_{1},S_{2},I}}\sum_{\vec{n}_{1},\vec{n}_{2}}G_{S_{1}S_{2}}%
(\vec{n}_{1}-\vec{n}_{2})\rho_{S_{1},I}^{a^{\dag},a}(\vec{n}_{1})\rho
_{S_{2},I}^{a^{\dag},a}(\vec{n}_{2})\right\}  \colon.
\end{align}
where $C$ is the coefficient of the Wigner SU(4)-invariant contact interaction
and $C_{I^{2}}$ is the coefficient of the isospin-dependent contact
interaction. $\ $For $C$ and $C_{I^{2}}$ we use the values%
\begin{equation}
C=\left(  3C^{I=1}+C^{I=0}\right)  /4, \label{C_coeff}%
\end{equation}%
\begin{equation}
C_{I^{2}}=\left(  C^{I=1}-C^{I=0}\right)  /4, \label{C_I2_coeff}%
\end{equation}
with $C^{I=1}=-5.021\times10^{-5}$ MeV$^{-2}$ and $C^{I=0}=-5.714\times
10^{-5}$ MeV$^{-2}$.

For the LO$_{2}$ transfer matrix we use the improved lattice transfer matrix
defined in \cite{Borasoy:2006qn},%
\begin{align}
M_{\text{LO}_{2}}  &  \equiv\colon\exp\left\{  -H_{\text{free}}\alpha
_{t}-\frac{\alpha_{t}}{2L^{3}}\sum_{\vec{q}}f(q^{2})\left[  C\rho^{a^{\dag}%
,a}(\vec{q})\rho^{a^{\dag},a}(-\vec{q})+C_{I^{2}}\sum_{I}\rho_{I}^{a^{\dag}%
,a}(\vec{q})\rho_{I}^{a^{\dag},a}(-\vec{q})\right]  \right. \nonumber\\
&  +\left.  \frac{g_{A}^{2}\alpha_{t}^{2}}{8f_{\pi}^{2}q_{\pi}}\sum
_{\substack{S_{1},S_{2},I}}\sum_{\vec{n}_{1},\vec{n}_{2}}G_{S_{1}S_{2}}%
(\vec{n}_{1}-\vec{n}_{2})\rho_{S_{1},I}^{a^{\dag},a}(\vec{n}_{1})\rho
_{S_{2},I}^{a^{\dag},a}(\vec{n}_{2})\right\}  \colon.
\end{align}
where the momentum-dependent coefficient function $f(q^{2})$ is defined as
\begin{equation}
f(q^{2})=f_{0}^{-1}\exp\left[  -b%
%TCIMACRO{\dsum \limits_{l}}%
%BeginExpansion
{\displaystyle\sum\limits_{l}}
%EndExpansion
\left(  1-\cos q_{l}\right)  \right]  ,
\end{equation}
and the normalization factor $f_{0}$ is determined by the condition%
\begin{equation}
f_{0}=\frac{1}{L^{3}}\sum_{\vec{q}}\exp\left[  -b%
%TCIMACRO{\dsum \limits_{l}}%
%BeginExpansion
{\displaystyle\sum\limits_{l}}
%EndExpansion
\left(  1-\cos q_{l}\right)  \right]  .
\end{equation}
As in \cite{Borasoy:2006qn} we use the value $b=0.6$. \ This gives
approximately the correct average effective range for the two $S$-wave
channels when $C$ and $C_{I^{2}}$ are properly tuned. \ For $C$ and $C_{I^{2}%
}$ we use $C^{I=1}=-3.414\times10^{-5}$ MeV$^{-2}$ and $C^{I=0}=-4.780\times
10^{-5}$ MeV$^{-2}$ and the relations in Eq.~(\ref{C_coeff}) and
(\ref{C_I2_coeff}). \ More details on these leading-order actions can be found
in \cite{Borasoy:2006qn}.

\subsection{Next-to-leading-order actions NLO$_{1}$ and NLO$_{2}$}

For the next-to-leading-order transfer matrices $M_{\text{NLO}_{1}}$ and
$M_{\text{NLO}_{2}}$ we add the following local interactions to the
leading-order transfer matrices $M_{\text{LO}_{1}}$ and $M_{\text{LO}_{2}}$.
\ We first start with corrections to the leading-order contact interactions.
\ These can be written as%
\begin{equation}
\Delta V=\frac{1}{2}\Delta C:\sum\limits_{\vec{n}}\rho^{a^{\dagger},a}(\vec
{n})\rho^{a^{\dagger},a}(\vec{n}):,
\end{equation}%
\begin{equation}
\Delta V_{I^{2}}=\frac{1}{2}\Delta C_{I^{2}}:\sum\limits_{\vec{n},I}\rho
_{I}^{a^{\dagger},a}(\vec{n})\rho_{I}^{a^{\dagger},a}(\vec{n}):.
\end{equation}
At next-to-leading order there are seven independent contact interactions with
two derivatives. \ These can be written as%
\begin{equation}
V_{q^{2}}=-\frac{1}{2}C_{q^{2}}:\sum\limits_{\vec{n},l}\rho^{a^{\dagger}%
,a}(\vec{n})\triangledown_{l}^{2}\rho^{a^{\dagger},a}(\vec{n}):,
\end{equation}%
\begin{equation}
V_{I^{2},q^{2}}=-\frac{1}{2}C_{I^{2},q^{2}}:\sum\limits_{\vec{n},I,l}\rho
_{I}^{a^{\dagger},a}(\vec{n})\triangledown_{l}^{2}\rho_{I}^{a^{\dagger}%
,a}(\vec{n}):,
\end{equation}%
\begin{equation}
V_{S^{2},q^{2}}=-\frac{1}{2}C_{S^{2},q^{2}}:\sum\limits_{\vec{n},S,l}\rho
_{S}^{a^{\dagger},a}(\vec{n})\triangledown_{l}^{2}\rho_{S}^{a^{\dagger}%
,a}(\vec{n}):,
\end{equation}%
\begin{equation}
V_{S^{2},I^{2},q^{2}}=-\frac{1}{2}C_{S^{2},I^{2},q^{2}}:\sum\limits_{\vec
{n},S,I,l}\rho_{S,I}^{a^{\dagger},a}(\vec{n})\triangledown_{l}^{2}\rho
_{S,I}^{a^{\dagger},a}(\vec{n}):,
\end{equation}%
\begin{equation}
V_{(q\cdot S)^{2}}=\frac{1}{2}C_{(q\cdot S)^{2}}:\sum\limits_{\vec{n}}%
\sum\limits_{S}\Delta_{S}\rho_{S}^{a^{\dagger},a}(\vec{n})\sum
\limits_{S^{\prime}}\Delta_{S^{\prime}}\rho_{S^{\prime}}^{a^{\dagger},a}%
(\vec{n}):,
\end{equation}%
\begin{equation}
V_{I^{2},(q\cdot S)^{2}}=\frac{1}{2}C_{I^{2},(q\cdot S)^{2}}:\sum
\limits_{\vec{n},I}\sum\limits_{S}\Delta_{S}\rho_{S,I}^{a^{\dagger},a}(\vec
{n})\sum\limits_{S^{\prime}}\Delta_{S^{\prime}}\rho_{S^{\prime},I}%
^{a^{\dagger},a}(\vec{n}):,
\end{equation}%
\begin{equation}
V_{(iq\times S)\cdot k}=-\frac{i}{2}C_{(iq\times S)\cdot k}:\sum
\limits_{\vec{n},l,S,l^{\prime}}\varepsilon_{l,S,l^{\prime}}\left[  \Pi
_{l}^{a^{\dagger},a}(\vec{n})\Delta_{l^{\prime}}\rho_{S}^{a^{\dagger},a}%
(\vec{n})+\Pi_{l,S}^{a^{\dagger},a}(\vec{n})\Delta_{l^{\prime}}\rho
^{a^{\dagger},a}(\vec{n})\right]  :.
\end{equation}
The subscripts indicate the continuum limit of the interactions. \ A detailed
discussion of the continuum limit for each of these NLO lattice interactions
is given in the appendix.

The $V_{(iq\times S)\cdot k}$ term corresponds with the continuum interaction%
\begin{equation}
C_{(iq\times S)\cdot k}\left(  i\vec{q}\times\left(  \vec{\sigma}_{1}%
+\vec{\sigma}_{2}\right)  \right)  \cdot\vec{k}.
\end{equation}
We note that this interaction vanishes unless the total spin $S=1$. \ We note
also that the continuum limit of the interaction is antisymmetric under the
exchange of $\vec{q}$ and $\vec{k}$. \ Therefore the continuum interaction is
nonzero only for odd parity channels. \ Unfortunately the lattice interaction
$V_{(iq\times S)\cdot k}$ does not share this exact $t$-$u$ channel
antisymmetry at nonzero lattice spacing. \ Therefore $V_{(iq\times S)\cdot k}$
produces small lattice artifacts for $S=1$ in even parity channels.
\ Fortunately in this case there is a simple way to remove them. \ We include
an explicit projection onto total isospin $I=1$,%
\begin{align}
V_{(iq\times S)\cdot k}^{I=1}  &  =-\frac{i}{2}C_{(iq\times S)\cdot k}%
^{I=1}\left\{  \frac{3}{4}:\sum\limits_{\vec{n},l,S,l^{\prime}}\varepsilon
_{l,S,l^{\prime}}\left[  \Pi_{l}^{a^{\dagger},a}(\vec{n})\Delta_{l^{\prime}%
}\rho_{S}^{a^{\dagger},a}(\vec{n})+\Pi_{l,S}^{a^{\dagger},a}(\vec{n}%
)\Delta_{l^{\prime}}\rho^{a^{\dagger},a}(\vec{n})\right]  :\right. \nonumber\\
&  +\left.  \frac{1}{4}:\sum\limits_{\vec{n},l,S,l^{\prime},I}\varepsilon
_{l,S,l^{\prime}}\left[  \Pi_{l,I}^{a^{\dagger},a}(\vec{n})\Delta_{l^{\prime}%
}\rho_{S,I}^{a^{\dagger},a}(\vec{n})+\Pi_{l,S,I}^{a^{\dagger},a}(\vec
{n})\Delta_{l^{\prime}}\rho_{I}^{a^{\dagger},a}(\vec{n})\right]  :\right\}  .
\end{align}
In the continuum limit the interaction is already pure $I=1$ due to total
antisymmetry, and so the isospin triplet projection has no effect. \ At
nonzero lattice spacing this projection completely eliminates lattice
artifacts in the $S=1$ even parity channels.

\section{Two-nucleon scattering on the lattice}

\subsection{Spherical wall method}

\begin{figure}
[ptb]
\begin{center}
\includegraphics[
height=1.6259in,
width=1.6259in
]%
{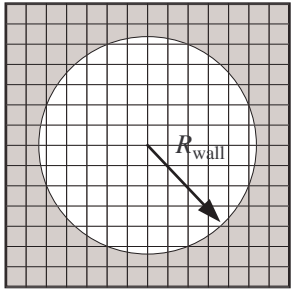}%
\caption{Spherical wall imposed in the center-of-mass frame.}%
\label{spherical_wall}%
\end{center}
\end{figure}

We measure phase shifts by imposing a hard spherical wall boundary on the
relative separation between the two nucleons at some chosen radius
$R_{\text{wall}}$ \cite{Borasoy:2007vy}. \ The spherical wall removes copies
of the interactions due to the periodic boundaries of the lattice. \ Viewed in
the center-of-mass frame we solve the Schr\"{o}dinger equation for spherical
standing waves which vanish at $r=R_{\text{wall}}$ as indicated in
Fig.~\ref{spherical_wall}.%
%TCIMACRO{\FRAME{ftbpFU}{1.6259in}{1.6259in}{0pt}{\Qcb{Spherical wall imposed
%in the center-of-mass frame.}}{\Qlb{spherical_wall}}{spherical_wall.eps}%
%width 1.6259in;  height 1.6259in;  depth 0pt;  original-width 3.1955in;
%original-height 3.1955in;  cropleft "0";  croptop "1";  cropright "1";
%cropbottom "0";  filename 'spherical_wall.eps';file-properties "XNPEU";}} }%
%BeginExpansion
%EndExpansion

In the spin singlet case for values of $r$ beyond the range of the
interaction, the spherical standing wave can be decomposed as a superposition
of products of spherical harmonics and spherical Bessel functions.
\ Explicitly we have%
\begin{equation}
\left[  \cos\delta_{L}\cdot j_{L}(kr)-\sin\delta_{L}\cdot y_{L}(kr)\right]
Y_{L,L_{z}}(\theta,\phi), \label{wavefunction}%
\end{equation}
where the center-of-mass energy of the spherical wave is%
\begin{equation}
E=2\frac{k^{2}}{2m}=\frac{k^{2}}{m},
\end{equation}
and the phase shift for partial wave $L$ is $\delta_{L}$. \ Therefore we know
$k$ from the energy $E$, and the phase shift $\delta_{L}$ is determined by
setting the wavefunction in Eq.~(\ref{wavefunction}) equal to zero at the wall
boundary,%
\begin{equation}
\cos\delta_{L}\cdot j_{L}(kR_{\text{wall}})=\sin\delta_{L}\cdot y_{L}%
(kR_{\text{wall}}),
\end{equation}%
\begin{equation}
\delta_{L}=\tan^{-1}\left[  \frac{j_{L}(kR_{\text{wall}})}{y_{L}%
(kR_{\text{wall}})}\right]  . \label{simple_phaseshift}%
\end{equation}
On the lattice there is some ambiguity on the precise value of $R_{\text{wall}%
}$ since the components of $\vec{r}$ must be integer multiples of the lattice
spacing. \ We resolve this ambiguity by fine-tuning the value of
$R_{\text{wall}}$ for each standing wave so that $\delta_{L}$ equals zero when
the particles are non-interacting.

For the spin triplet case, however, spin-orbit coupling produces mixing
between partial waves $L=J-1$ and $L=J+1.$ \ If we use the two-component
notation,
\begin{equation}
\left[
\begin{array}
[c]{c}%
R_{J-1}(r)\\
R_{J+1}(r)
\end{array}
\right]  , \label{twocomponent}%
\end{equation}
for the radial part of the wavefunction, then we find two sets of standing
wave solutions of the form%
\begin{equation}
\Psi^{I}\propto\frac{1}{k^{I}r}\left[
\begin{array}
[c]{c}%
A_{J-1}^{I}\sin\left(  k^{I}r-\frac{J-1}{2}\pi+\Delta_{J-1}^{I}\right) \\
A_{J+1}^{I}\sin\left(  k^{I}r-\frac{J+1}{2}\pi+\Delta_{J+1}^{I}\right)
\end{array}
\right]
\end{equation}
at energy $E^{I}=(k^{I})^{2}/m$ and%
\begin{equation}
\Psi^{II}\propto\frac{1}{k^{II}r}\left[
\begin{array}
[c]{c}%
A_{J-1}^{II}\sin\left(  k^{II}r-\frac{J-1}{2}\pi+\Delta_{J-1}^{II}\right) \\
A_{J+1}^{II}\sin\left(  k^{II}r-\frac{J+1}{2}\pi+\Delta_{J+1}^{II}\right)
\end{array}
\right]
\end{equation}
at $E^{II}=(k^{II})^{2}/m$. \ These can be used to derive the phase shifts
$\delta_{J-1}$ and $\delta_{J+1}$ and mixing angle $\varepsilon_{J}$
\cite{Borasoy:2007vy},%
\begin{equation}
\tan\left(  -\Delta_{J-1}^{I}+\delta_{J-1}\right)  \tan\left(  -\Delta
_{J+1}^{I}+\delta_{J+1}\right)  =\tan^{2}\varepsilon_{J},
\label{newconstraint1}%
\end{equation}%
\begin{equation}
\tan\left(  -\Delta_{J-1}^{II}+\delta_{J-1}\right)  \tan\left(  -\Delta
_{J+1}^{II}+\delta_{J+1}\right)  =\tan^{2}\varepsilon_{J},
\label{newconstraint2}%
\end{equation}%
\begin{equation}
A_{J-1}^{I}\tan\varepsilon_{J}=-A_{J+1}^{I}\frac{\sin\left(  -\Delta_{J+1}%
^{I}+\delta_{J+1}\right)  }{\cos\left(  -\Delta_{J-1}^{I}+\delta_{J-1}\right)
}, \label{newconstraint3}%
\end{equation}%
\begin{equation}
A_{J-1}^{II}\tan\varepsilon_{J}=-A_{J+1}^{II}\frac{\sin\left(  -\Delta
_{J+1}^{II}+\delta_{J+1}\right)  }{\cos\left(  -\Delta_{J-1}^{II}+\delta
_{J-1}\right)  }. \label{newconstraint4}%
\end{equation}

The phase shifts and mixing angle in Eq.~(\ref{newconstraint1}) and
(\ref{newconstraint3}) are at momentum $k^{I}$ while the phase shifts and
mixing angle in Eq.~(\ref{newconstraint2}) and (\ref{newconstraint4}) are at
momentum $k^{II}$. \ We therefore consider only close pairs of values
$k^{I}\approx k^{II}$ in solving Eq.~(\ref{newconstraint1}%
)-(\ref{newconstraint4}). \ This can be done for example by considering the
$(n+1)^{\text{st}}$-radial excitation of $L=J-1$ together with the
$n^{\text{th}}$-radial excitation of $L=J+1$. \ In this scheme we use%
\begin{equation}
\tan\left(  -\Delta_{J-1}^{I}+\delta_{J-1}(k^{I})\right)  \tan\left(
-\Delta_{J+1}^{I}+\delta_{J+1}(k^{I})\right)  =\tan^{2}\left[  \varepsilon
_{J}(k^{I})\right]  , \label{kI_1}%
\end{equation}%
\begin{equation}
\tan\left(  -\Delta_{J-1}^{II}+\delta_{J-1}(k^{I})\right)  \tan\left(
-\Delta_{J+1}^{II}+\delta_{J+1}(k^{I})\right)  \approx\tan^{2}\left[
\varepsilon_{J}(k^{I})\right]  , \label{kI_2}%
\end{equation}%
\begin{equation}
A_{J-1}^{I}\tan\left[  \varepsilon_{J}(k^{I})\right]  =-A_{J+1}^{I}\frac
{\sin\left(  -\Delta_{J+1}^{I}+\delta_{J+1}(k^{I})\right)  }{\cos\left(
-\Delta_{J-1}^{I}+\delta_{J-1}(k^{I})\right)  }, \label{kI_3}%
\end{equation}
for the phase shifts and mixing angle at $k=k^{I}$, and%
\begin{equation}
\tan\left(  -\Delta_{J-1}^{I}+\delta_{J-1}(k^{II})\right)  \tan\left(
-\Delta_{J+1}^{I}+\delta_{J+1}(k^{II})\right)  \approx\tan^{2}\left[
\varepsilon_{J}(k^{II})\right]  , \label{kII_1}%
\end{equation}%
\begin{equation}
\tan\left(  -\Delta_{J-1}^{II}+\delta_{J-1}(k^{II})\right)  \tan\left(
-\Delta_{J+1}^{II}+\delta_{J+1}(k^{II})\right)  =\tan^{2}\left[
\varepsilon_{J}(k^{II})\right]  , \label{kII_2}%
\end{equation}%
\begin{equation}
A_{J-1}^{II}\tan\left[  \varepsilon_{J}(k^{II})\right]  =-A_{J+1}^{II}%
\frac{\sin\left(  -\Delta_{J+1}^{II}+\delta_{J+1}(k^{II})\right)  }%
{\cos\left(  -\Delta_{J-1}^{II}+\delta_{J-1}(k^{II})\right)  }, \label{kII_3}%
\end{equation}
for the phase shifts and mixing angle at $k=k^{II}$.

For momentum less than or equal to the pion mass all of the mixing angles
$\varepsilon_{J}$ are numerically small. \ It is therefore convenient to
expand in powers of the mixing angle,%
\begin{equation}
\delta_{J-1}(k^{I})=\Delta_{J-1}^{I}+\frac{\varepsilon_{J}^{2}(k^{I})}%
{\tan\left(  -\Delta_{J+1}^{I}+\delta_{J+1}(k^{I})\right)  }+O(\varepsilon
_{J}^{4}),
\end{equation}%
\begin{equation}
\varepsilon_{J}(k^{I})=-\frac{A_{J+1}^{I}}{A_{J-1}^{I}}\sin\left(
\Delta_{J+1}^{II}-\Delta_{J+1}^{I}\right)  +O(\varepsilon_{J}^{3}),
\end{equation}
at $k=k^{I}$ and%
\begin{equation}
\delta_{J+1}(k^{II})=\Delta_{J+1}^{II}+\frac{\varepsilon_{J}^{2}(k^{II})}%
{\tan\left(  -\Delta_{J-1}^{II}+\delta_{J-1}(k^{II})\right)  }+O(\varepsilon
_{J}^{4}),
\end{equation}%
\begin{equation}
\varepsilon_{J}(k^{II})=\frac{A_{J-1}^{II}}{A_{J+1}^{II}}\sin\left(
\Delta_{J-1}^{II}-\Delta_{J-1}^{I}\right)  +O(\varepsilon_{J}^{3}),
\end{equation}
at $k=k^{II}$.

\subsection{LO energy levels and NLO corrections}

\begin{figure}
[ptb]
\begin{center}
\includegraphics[
height=3.7749in,
width=4.6959in
]%
{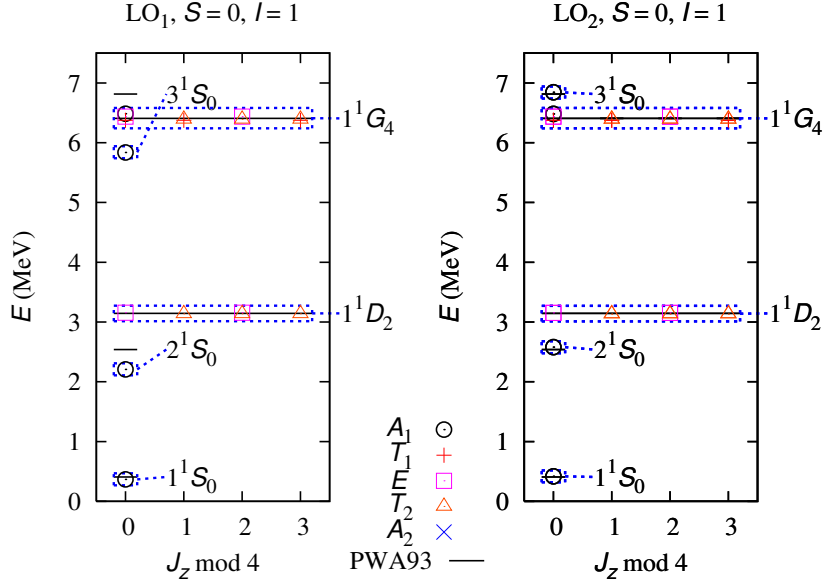}%
\caption{Energy levels for $S=0$, $I=1$ using lattice actions LO$_{1}$ and
LO$_{2}$ and a spherical wall at radius $R_{\text{wall}}=10+\epsilon$ lattice
units. \ The solid line indicates the exact energy levels which reproduce data
from the partial wave analysis of \cite{Stoks:1993tb}.}%
\label{s0_i1_r10}%
\end{center}
\end{figure}

In Fig.~\ref{s0_i1_r10} we show energy levels for spin $S=0$ and isospin $I=1$
using lattice actions LO$_{1}$ and LO$_{2}$. \ The spherical wall is at radius
$R_{\text{wall}}=10+\epsilon$ lattice units where $\epsilon$ is a small
positive number. \ We use this $\epsilon$ notation to make explicit that
$\left\vert \vec{r}\right\vert =10$ lattice units\ is inside the spherical
wall but all lattice sites with $\left\vert \vec{r}\right\vert >10$ lattice
units lie outside. \ The solid lines indicate the exact energy levels which
reproduce data from the partial wave analysis of \cite{Stoks:1993tb}.%
%TCIMACRO{\FRAME{ftbpFU}{4.6959in}{3.7749in}{0pt}{\Qcb{Energy levels for $S=0$,
%$I=1$ using lattice actions LO$_{1}$ and LO$_{2}$ and a spherical wall at
%radius $R_{\text{wall}}=10+\epsilon$ lattice units. \ The solid line indicates
%the exact energy levels which reproduce data from the partial wave analysis of
%\cite{Stoks:1993tb}.}}{\Qlb{s0_i1_r10}}{s0_i1_r10.eps}%
%width 4.6959in;  height 3.7749in;  depth 0pt;  original-width 7.7798in;
%original-height 6.2457in;  cropleft "0";  croptop "1";  cropright "1";
%cropbottom "0";  filename 'S0_I1_R10.eps';file-properties "XNPEU";}} }%
%BeginExpansion
%EndExpansion
The energy levels for the standard action LO$_{1}$ are about $10\%$ to $15\%$
too low for the $^{1}S_{0}$ states, while the improved action LO$_{2}$ is
correct to within a couple of percent for all $^{1}S_{0}$ states$.$
\ Deviations for the higher partial waves are smaller than $1\%$ for both
LO$_{1}$ and LO$_{2}$.

\begin{figure}
[ptb]
\begin{center}
\includegraphics[
height=3.7749in,
width=4.6959in
]%
{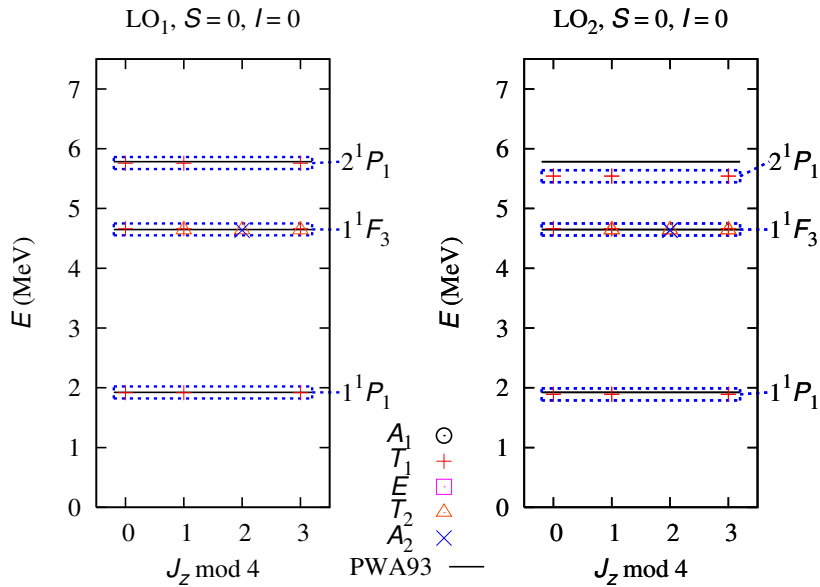}%
\caption{Energy levels for $S=0$, $I=0$ using lattice actions LO$_{1}$ and
LO$_{2}$ and a spherical wall at radius $R_{\text{wall}}=10+\epsilon$ lattice
units. \ The solid line indicates the exact energy levels which reproduce data
from the partial wave analysis of \cite{Stoks:1993tb}.}%
\label{s0_i0_r10}%
\end{center}
\end{figure}

The energy levels for spin $S=0$, isospin $I=0$, and $R_{\text{wall}%
}=10+\epsilon$ lattice units are shown in Fig.~\ref{s0_i0_r10}.%
%TCIMACRO{\FRAME{ftbpFU}{4.6959in}{3.7749in}{0pt}{\Qcb{Energy levels for $S=0$,
%$I=0$ using lattice actions LO$_{1}$ and LO$_{2}$ and a spherical wall at
%radius $R_{\text{wall}}=10+\epsilon$ lattice units. \ The solid line indicates
%the exact energy levels which reproduce data from the partial wave analysis of
%\cite{Stoks:1993tb}.}}{\Qlb{s0_i0_r10}}{s0_i0_r10.eps}%
%width 4.6959in;  height 3.7749in;  depth 0pt;  original-width 6.992in;
%original-height 10.0024in;  cropleft "0";  croptop "1";  cropright "1";
%cropbottom "0";  filename 'S0_I0_R10.eps';file-properties "XNPEU";}} }%
%BeginExpansion
%EndExpansion
In this case LO$_{1}$ is better for the $^{1}P_{1}$ states and is within $1\%$
of the exact values. \ The LO$_{2}$ energy levels are further away, though
still within $5\%$ for the $^{1}P_{1}$ states. \ The error for the $^{1}F_{3}$
partial wave is much smaller than $1\%$ for both LO$_{1}$ and LO$_{2}$.
\ Results for the LO$_{1}$ and LO$_{2}$ in the spin-triplet channels are
similar in terms of their relative errors.

At next-to-leading order\ we have nine unknown operator coefficients to fit:
$\ \Delta C$, $\Delta C_{I^{2}},$ $C_{q^{2}},$ $C_{I^{2},q^{2}},$
$C_{S^{2},q^{2}}$, $C_{S^{2},I^{2},q^{2}}$, $C_{(q\cdot S)^{2}}$,
$C_{I^{2},(q\cdot S)^{2}}$, and $C_{(iq\times S)\cdot k}^{I=1}$. \ We fit
these nine operator coefficients using the eight energy levels listed in Table
\ref{fitvalues} for $R_{\text{wall}}=10+\epsilon$ lattice units, as well as
the quadrupole moment of the deuteron $Q_{d}$. \ The deuteron quadrupole
moment is a measure of the low-energy strength of $S$-$D$ partial wave mixing.
\ We could instead use\ the mixing angle $\varepsilon_{1}$ at some low
momentum scale, but the quadrupole moment is actually an easier observable to
compute. \ \begin{table}[tb]
\caption{Results for LO$_{1}$ and LO$_{2}$ and target values}%
\label{fitvalues}
\begin{tabular}
[c]{||c|c|c|c|c||}\hline\hline
Spherical wave & Free nucleons & LO$_{1}$ & LO$_{2}$ & PWA93\\\hline
$1^{1}S_{0}$ (MeV) & $0.928$ & $0.368$ & $0.418$ & $0.407$\\\hline
$3^{1}S_{0}$ (MeV) & $8.535$ & $5.838$ & $6.843$ & $6.815$\\\hline
$1^{3}S(D)_{1}$ (MeV) & $0.928$ & $-2.225$ & $-2.225$ & $-2.225$\\\hline
$3^{3}S(D)_{1}$ (MeV) & $8.535$ & $4.878$ & $5.430$ & $5.675$\\\hline
$2^{1}P_{1}$ (MeV) & $5.691$ & $5.755$ & $5.541$ & $5.782$\\\hline
$2^{3}P(F)_{0}$ (MeV) & $5.691$ & $5.569$ & $5.396$ & $5.584$\\\hline
$2^{3}P(F)_{1}$ (MeV) & $5.691$ & $5.754$ & $5.652$ & $5.753$\\\hline
$2^{3}P(F)_{2}$ (MeV) & $5.691$ & $5.684$ & $5.558$ & $5.669$\\\hline
$Q_{d}$ (fm$^{2}$) & N/A & $0.143$ & $0.276$ & $0.286$\\\hline\hline
\end{tabular}
\end{table}In this study we incorporate the nine NLO interactions using
perturbation theory.

The motivation for using perturbation theory for the NLO\ interactions is that
it is computationally less expensive. Whether or not the NLO corrections can
be treated using perturbation theory depends on the observable of interest.
For example the standard action LO$_{1}$ has a clustering instability which
results in a strong overbinding of the alpha particle \cite{Borasoy:2006qn}.
\ This instability is severe and probably cannot be fixed by introducing NLO
corrections perturbatively. \ The problem is that the $S$-wave interactions
are too attractive at momenta $q\sim m_{\pi}$, and the $S$-wave effective
range corrections provided at NLO must be included nonperturbatively.

On the other hand the improved action LO$_{2}$ resolves this problem using
Gaussian smearing. \ The Gaussian-smeared interaction is contained in the term
$V_{2}^{Q^{2}/\Lambda^{2}}$ defined in Eq.~(\ref{LO2}). \ Since $V_{2}%
^{Q^{2}/\Lambda^{2}}$ is iterated nonperturbatively along with rest of the
leading-order action, the clustering instability is removed. \ We use this
approach as a general strategy. \ The first attempt is to try a completely
perturbative approach for all NLO corrections. \ If this fails and some of the
interactions must be handled nonperturbatively, then we include those
interactions in the $V^{Q^{2}/\Lambda^{2}}$ term in the leading-order action.

For each of the nine observables in Table \ref{fitvalues} we compute the
derivative with respect to each of the nine NLO coefficient operators. \ By
inverting this $9\times9$ Jacobian matrix we find the required values for the
operator coefficients needed to match each of the observables using
first-order perturbation theory. \ The results for the operator coefficients
are shown in Table \ref{coefficients}. \ Although most of coefficients have
generally the same order of magnitude and sign for NLO$_{1}$ and NLO$_{2}$, we
see that the coefficients in some cases are quite different.\begin{table}[tb]
\caption{Results for NLO operator coefficients}%
\label{coefficients}
\begin{tabular}
[c]{||c|c|c||}\hline\hline
Coefficient & $\text{NLO}_{1}$ & $\text{NLO}_{2}$\\\hline
$\Delta C$ (MeV$^{-2}$) & $-1.43\times10^{-4}$ & $-1.10\times10^{-5}$\\\hline
$\Delta C_{I^{2}}$ (MeV$^{-2}$) & $3.42\times10^{-5}$ & $1.03\times10^{-5}%
$\\\hline
$C_{q^{2}}$ (MeV$^{-4}$) & $1.27\times10^{-9}$ & $-1.01\times10^{-9}$\\\hline
$C_{I^{2},q^{2}}$ (MeV$^{-4}$) & $-5.83\times10^{-10}$ & $-4.07\times10^{-10}%
$\\\hline
$C_{S^{2},q^{2}}$ (MeV$^{-4}$) & $-4.18\times10^{-12}$ & $-1.72\times10^{-10}%
$\\\hline
$C_{S^{2},I^{2},q^{2}}$ (MeV$^{-4}$) & $-4.31\times10^{-10}$ & $-2.81\times
10^{-10}$\\\hline
$C_{(q\cdot S)^{2}}$ (MeV$^{-4}$) & $-2.52\times10^{-11}$ & $-3.23\times
10^{-10}$\\\hline
$C_{I^{2},(q\cdot S)^{2}}$ (MeV$^{-4}$) & $6.24\times10^{-11}$ &
$1.43\times10^{-10}$\\\hline
$C_{(iq\times S)\cdot k}^{I=1}$ (MeV$^{-4}$) & $1.60\times10^{-10}$ &
$9.81\times10^{-11}$\\\hline\hline
\end{tabular}
\end{table}

\section{Results}

\begin{figure}
[ptb]
\begin{center}
\includegraphics[
height=2.6057in,
width=5.0289in
]%
{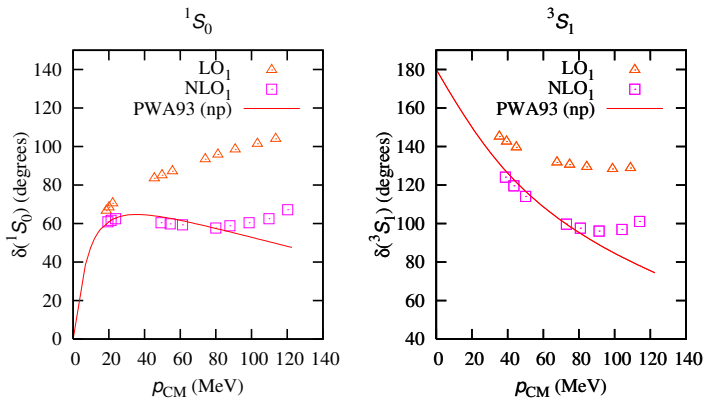}%
\caption{$S$-wave phase shifts versus center-of-mass momentum for LO$_{1}$ and
NLO$_{1}$.}%
\label{swave_b0}%
\end{center}
\end{figure}

\begin{figure}
[ptbptb]
\begin{center}
\includegraphics[
height=2.6048in,
width=5.0298in
]%
{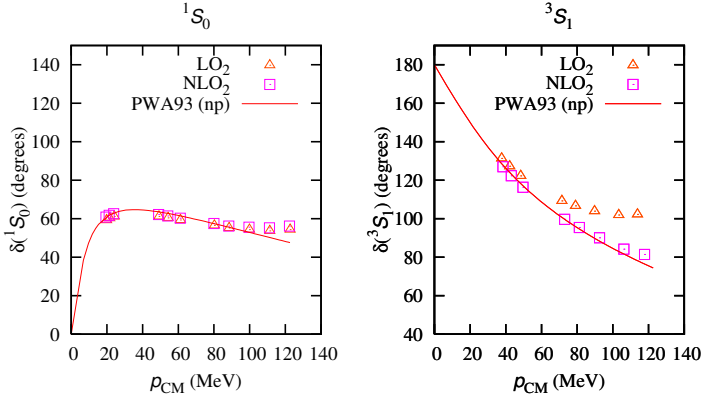}%
\caption{$S$-wave phase shifts versus center-of-mass momentum for LO$_{2}$ and
NLO$_{2}$.}%
\label{swave_b6}%
\end{center}
\end{figure}

We compute lattice phase shifts and mixing angles using spherical walls with
radii $R_{\text{wall}}=10+\epsilon$, $9+\epsilon$, and $8+\epsilon$ lattice
units. \ In order of increasing momentum, the lattice data corresponds with
the first radial excitation for $R_{\text{wall}}=10+\epsilon,9+\epsilon,$ and
$8+\epsilon$; second radial excitation of $R_{\text{wall}}=10+\epsilon
,9+\epsilon,$ and $8+\epsilon;$ and so on. \ The $S$-wave phase shifts for
LO$_{1}$ and NLO$_{1}$ versus center-of-mass momentum $p_{\text{CM}}$ are
shown in Fig.~\ref{swave_b0}. \ We compare these with the $S$-wave phase
shifts for LO$_{2}$ and NLO$_{2}$ in Fig.~\ref{swave_b6}. \ The NLO$_{1}$ and
NLO$_{2}$ results are both in good agreement with partial wave results from
\cite{Stoks:1993tb}. \ Systematic errors are first noticeable at momenta
greater than about $80$ MeV and are larger for NLO$_{1}$. \ In both cases the
deviations are consistent with higher-order effects such as two-nucleon
contact interactions with four derivatives.%
%TCIMACRO{\FRAME{ftbpFU}{5.0289in}{2.6057in}{0pt}{\Qcb{$S$-wave phase shifts
%versus center-of-mass momentum for LO$_{1}$ and NLO$_{1}$.}}{\Qlb{swave_b0}%
%}{swave_b0.eps}{\special{ language "Scientific Word";  type "GRAPHIC";
%width 5.0289in;  height 2.6057in;  depth 0pt;  original-width 2.4414in;
%original-height 3in;  cropleft "0";  croptop "1";  cropright "1";
%cropbottom "0";  filename 'Swave_b0.eps';file-properties "XNPEU";}} }%
%BeginExpansion
%EndExpansion%
%TCIMACRO{\FRAME{ftbpFU}{5.0298in}{2.6048in}{0pt}{\Qcb{$S$-wave phase shifts
%versus center-of-mass momentum for LO$_{2}$ and NLO$_{2}$.}}{\Qlb{swave_b6}%
%}{swave_b6.eps}{\special{ language "Scientific Word";  type "GRAPHIC";
%width 5.0298in;  height 2.6048in;  depth 0pt;  original-width 8.3359in;
%original-height 4.2964in;  cropleft "0";  croptop "1";  cropright "1";
%cropbottom "0";  filename 'Swave_b6.eps';file-properties "XNPEU";}} }%
%BeginExpansion
%EndExpansion

\begin{figure}
[ptb]
\begin{center}
\includegraphics[
height=5.1145in,
width=5.028in
]%
{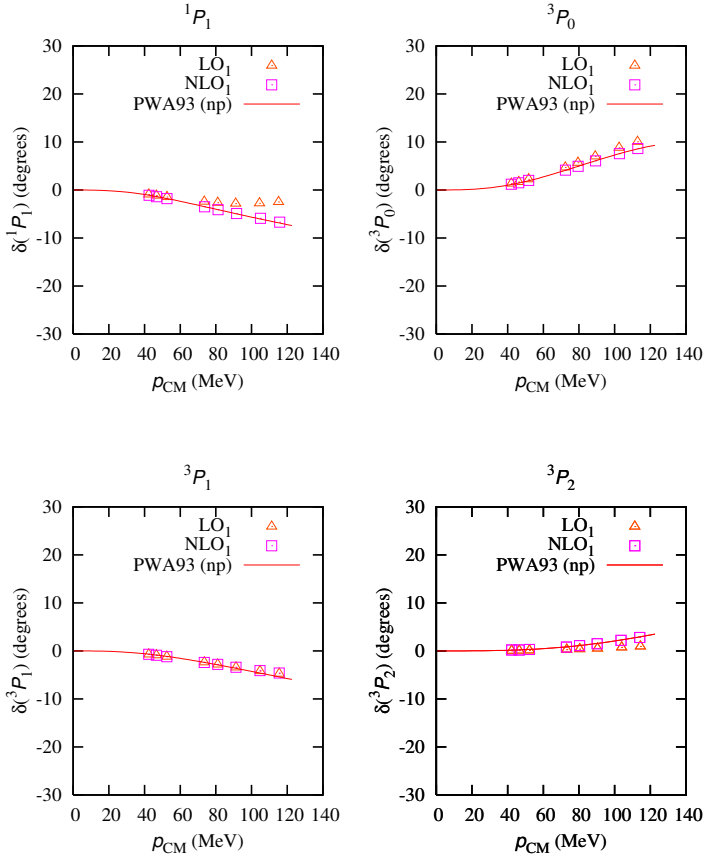}%
\caption{$P$-wave phase shifts versus center-of-mass momentum for LO$_{1}$ and
NLO$_{1}$.}%
\label{pwave_b0}%
\end{center}
\end{figure}

\begin{figure}
[ptbptb]
\begin{center}
\includegraphics[
height=5.1145in,
width=5.028in
]%
{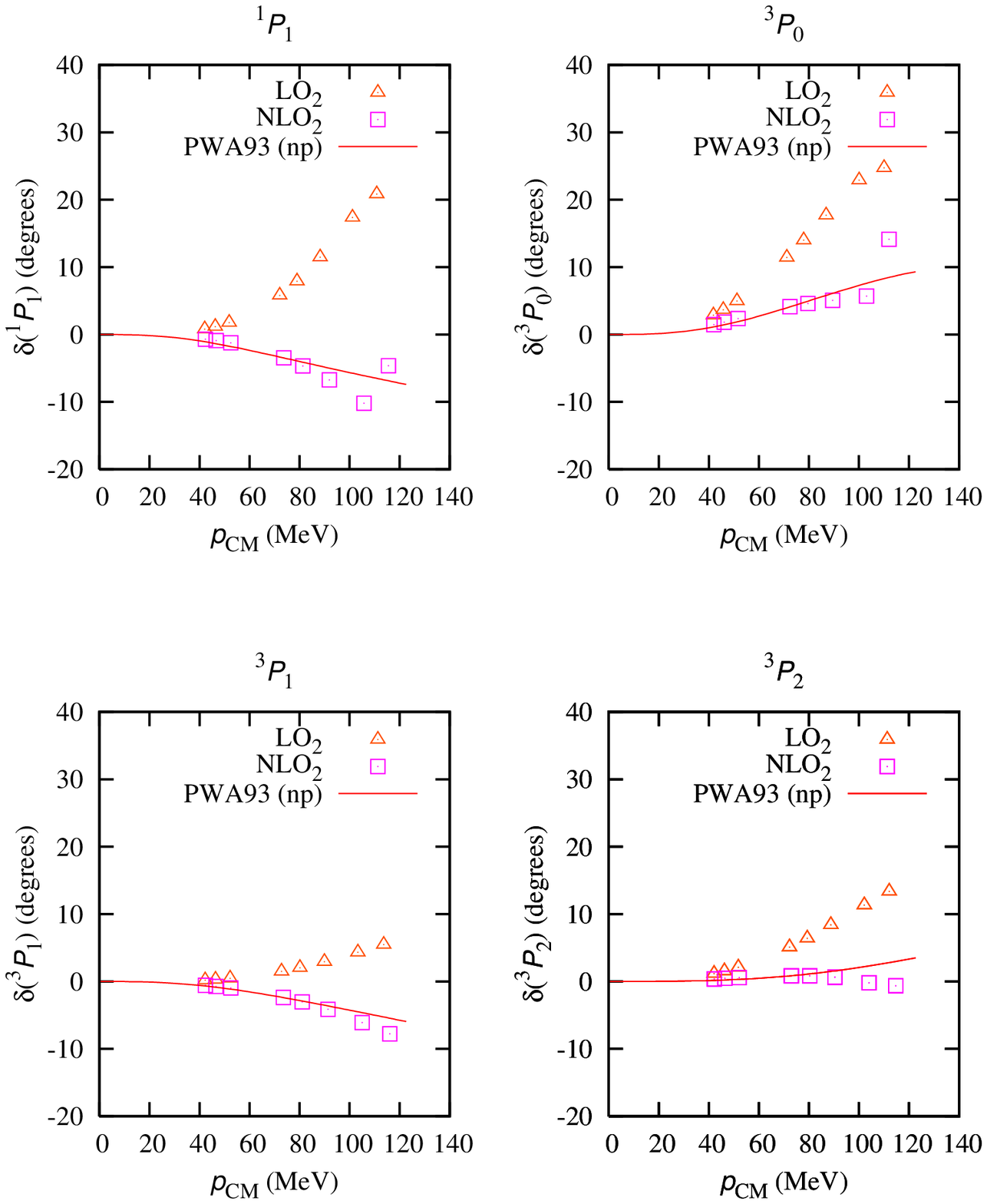}%
\caption{$P$-wave phase shifts versus center-of-mass momentum for LO$_{2}$ and
NLO$_{2}$.}%
\label{pwave_b6}%
\end{center}
\end{figure}

The $P$-wave phase shifts are presented in Fig.~\ref{pwave_b0} and
\ref{pwave_b6}. \ The phase shifts are already not bad for LO$_{1}$ and are
quite accurate for NLO$_{1}$. \ This indicates that at low momenta only a
small correction is needed on top of $P$-wave interactions produced by
one-pion exchange. \ In the case of LO$_{2}$ we see that the effect of
Gaussian smearing, while useful for $S$-wave phase shifts, produces an
unphysical attraction in each $P$-wave channel that must be cancelled by the
NLO$_{2}$ corrections. \ For the NLO$_{2}$ results the residual deviations
appear consistent with effects produced by higher-order terms such as
four-derivative contact interactions. \ For $^{1}P_{1}$ and $^{3}P_{0}$ there
is some indication of higher-derivative effects for $p_{\text{CM}}$ near $110$
MeV.%
%TCIMACRO{\FRAME{ftbpFU}{5.028in}{5.1145in}{0pt}{\Qcb{$P$-wave phase shifts
%versus center-of-mass momentum for LO$_{1}$ and NLO$_{1}$.}}{\Qlb{pwave_b0}%
%}{pwave_b0.eps}{\special{ language "Scientific Word";  type "GRAPHIC";
%width 5.028in;  height 5.1145in;  depth 0pt;  original-width 2.4414in;
%original-height 3in;  cropleft "0";  croptop "1";  cropright "1";
%cropbottom "0";  filename 'Pwave_b0.eps';file-properties "XNPEU";}} }%
%BeginExpansion
%EndExpansion%
%TCIMACRO{\FRAME{ftbpFU}{5.028in}{5.1145in}{0pt}{\Qcb{$P$-wave phase shifts
%versus center-of-mass momentum for LO$_{2}$ and NLO$_{2}$.}}{\Qlb{pwave_b6}%
%}{pwave_b6.eps}{\special{ language "Scientific Word";  type "GRAPHIC";
%width 5.028in;  height 5.1145in;  depth 0pt;  original-width 8.3333in;
%original-height 8.4769in;  cropleft "0";  croptop "1";  cropright "1";
%cropbottom "0";  filename 'Pwave_b6.eps';file-properties "XNPEU";}} }%
%BeginExpansion
%EndExpansion

\begin{figure}
[ptb]
\begin{center}
\includegraphics[
height=5.1136in,
width=5.0272in
]%
{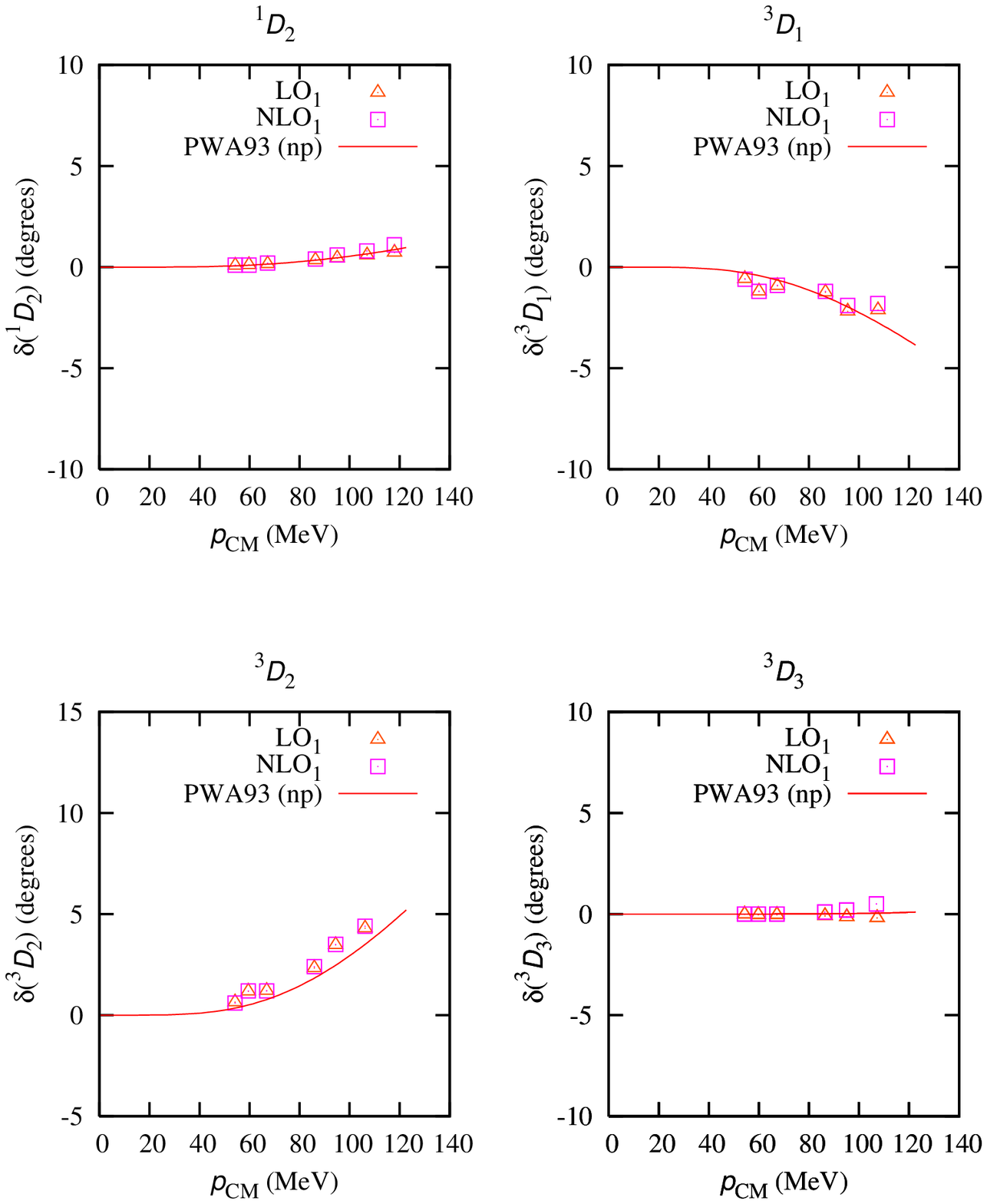}%
\caption{$D$-wave phase shifts versus center-of-mass momentum for LO$_{1}$ and
NLO$_{1}$.}%
\label{dwave_b0}%
\end{center}
\end{figure}

\begin{figure}
[ptbptb]
\begin{center}
\includegraphics[
height=5.1145in,
width=5.028in
]%
{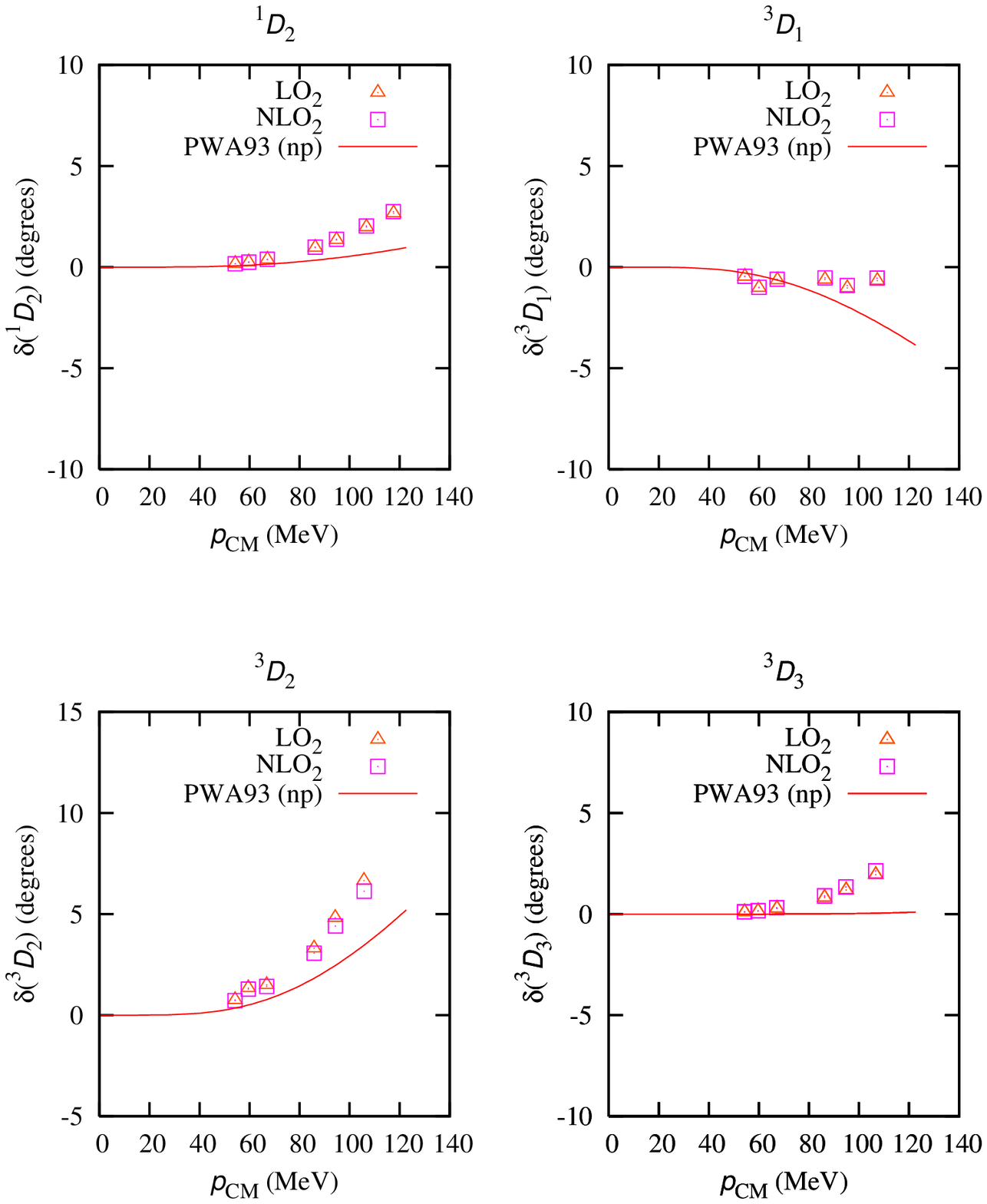}%
\caption{$D$-wave phase shifts versus center-of-mass momentum for LO$_{2}$ and
NLO$_{2}$.}%
\label{dwave_b6}%
\end{center}
\end{figure}

The $D$-wave phase shifts are shown in Fig.~\ref{dwave_b0} and \ref{dwave_b6}.
\ None of the $D$-wave data was used in the fitting of operator coefficients.
\ For both NLO$_{1}$ and NLO$_{2}$ results the errors appear consistent with
effects from higher-order interactions. \ The NLO$_{1}$ deviations are
somewhat smaller, though the NLO$_{1}$ and NLO$_{2}$ deviations appear similar
in character. \ Overall the differences among LO$_{1}$, LO$_{2}$, NLO$_{1}$,
and NLO$_{2}$ results for the $D$ waves are smaller than the corresponding
differences for the $S$ and $P$ waves. \ This observation is consistent with
the dominance of the one-pion exchange potential and validity of the Born
approximation in higher partial waves.%

%TCIMACRO{\FRAME{ftbpFU}{5.0272in}{5.1136in}{0pt}{\Qcb{$D$-wave phase shifts
%versus center-of-mass momentum for LO$_{1}$ and NLO$_{1}$.}}{\Qlb{dwave_b0}%
%}{dwave_b0.eps}{\special{ language "Scientific Word";  type "GRAPHIC";
%width 5.0272in;  height 5.1136in;  depth 0pt;  original-width 2.4414in;
%original-height 3in;  cropleft "0";  croptop "1";  cropright "1";
%cropbottom "0";  filename 'Dwave_b0.eps';file-properties "XNPEU";}} }%
%BeginExpansion
%EndExpansion%
%TCIMACRO{\FRAME{ftbpFU}{5.028in}{5.1145in}{0pt}{\Qcb{$D$-wave phase shifts
%versus center-of-mass momentum for LO$_{2}$ and NLO$_{2}$.}}{\Qlb{dwave_b6}%
%}{dwave_b6.eps}{\special{ language "Scientific Word";  type "GRAPHIC";
%width 5.028in;  height 5.1145in;  depth 0pt;  original-width 8.3333in;
%original-height 8.4769in;  cropleft "0";  croptop "1";  cropright "1";
%cropbottom "0";  filename 'Dwave_b6.eps';file-properties "XNPEU";}} }%
%BeginExpansion
%EndExpansion

\begin{figure}
[ptb]
\begin{center}
\includegraphics[
height=2.6048in,
width=5.0289in
]%
{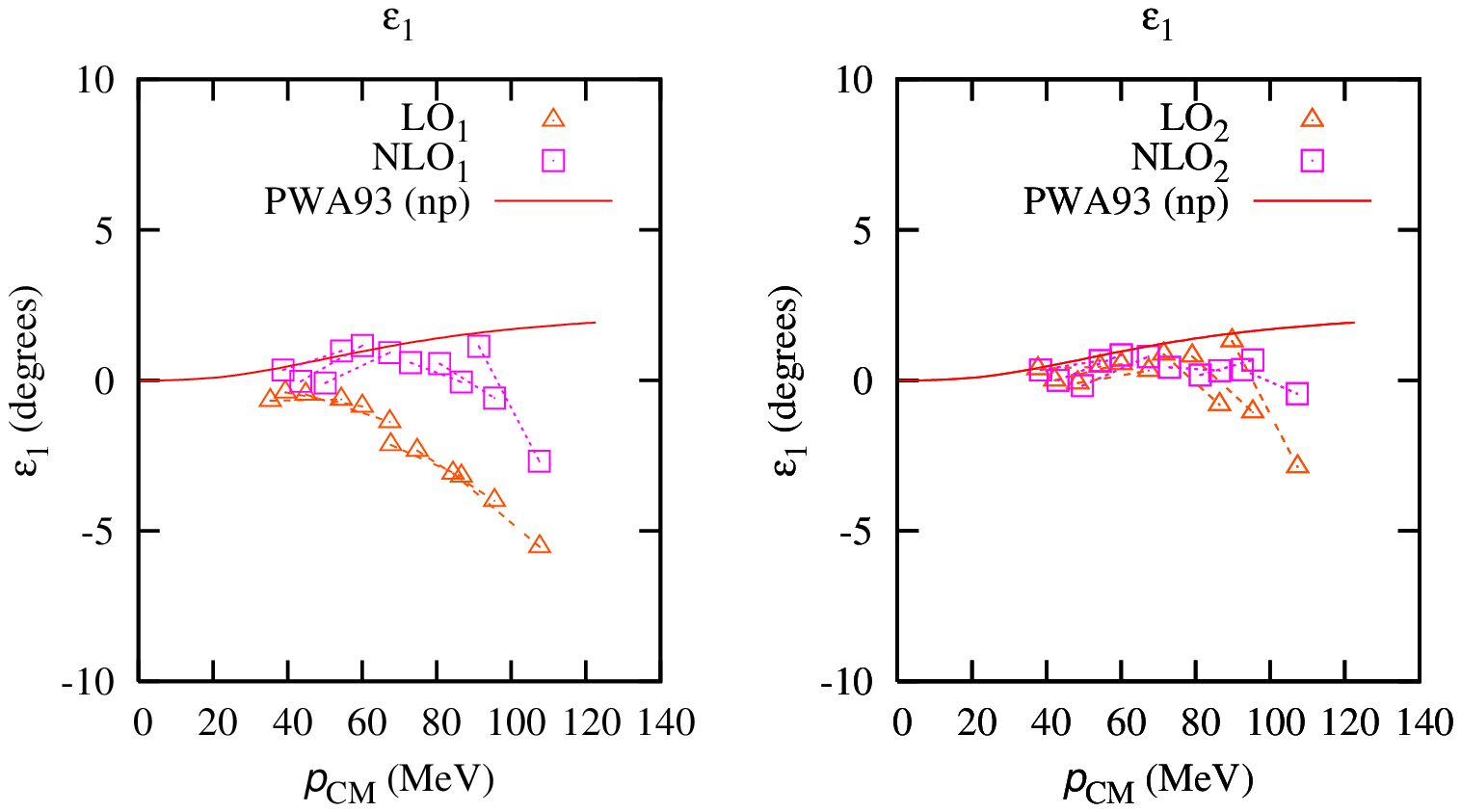}%
\caption{$\varepsilon_{1}$ mixing angle for LO$_{1}$ and NLO$_{1}$ on the
left, LO$_{2}$ and NLO$_{2}$ on the right.}%
\label{eps1}%
\end{center}
\end{figure}

The mixing parameter $\varepsilon_{1}$ in the Stapp parameterization
\cite{Stapp:1956mz} is shown in Fig.~\ref{eps1}. Results for LO$_{1}$ and
NLO$_{1}$ are on the left, and results for LO$_{2}$ and NLO$_{2}$ are on the
right. \ The pairs of points connected by dotted lines indicate pairs of
solutions at $k=k^{I}$ and $k=k^{II}$ for the coupled $^{3}S_{1}$-$^{3}D_{1}$
channels.\ \ For LO$_{1}$ it is interesting to note that $\varepsilon_{1}$ has
the wrong sign. \ For both NLO$_{1}$ and NLO$_{2}$ results the remaining
deviations appear consistent with effects produced by higher-order
interactions.%
%TCIMACRO{\FRAME{ftbpFU}{5.0289in}{2.6048in}{0pt}{\Qcb{$\varepsilon_{1}$ mixing
%angle for LO$_{1}$ and NLO$_{1}$ on the left, LO$_{2}$ and NLO$_{2}$ on the
%right.}}{\Qlb{eps1}}{eps1.eps}{\special{ language "Scientific Word";
%type "GRAPHIC";  maintain-aspect-ratio TRUE;  display "USEDEF";
%valid_file "F";  width 5.0289in;  height 2.6048in;  depth 0pt;
%original-width 2.4414in;  original-height 3in;  cropleft "0";  croptop "1";
%cropright "1";  cropbottom "0";  filename 'eps1.eps';file-properties "XNPEU";}%
%} }%
%BeginExpansion
%EndExpansion

\section{Summary and discussion}

We have studied nucleon-nucleon scattering on the lattice at next-to-leading
order in chiral effective field theory at momenta less than or equal to the
pion mass. \ Throughout our analysis we tested model independence at fixed
lattice spacing by repeating calculations using two different lattice actions.
\ The first leading-order action LO$_{1}$ included instantaneous one-pion
exchange and same-site contact interactions. \ The second leading-order action
LO$_{2}$ included instantaneous one-pion exchange and Gaussian-smeared
interactions. \ We computed next-to-leading-order corrections for these
actions and in each case found results accurate up to corrections at higher
order. \ In the second paper of this series we use the LO$_{2}$ and NLO$_{2}$
actions to compute the ground state of dilute neutron matter using Monte
Carlo. \ This is done using the auxiliary-field transfer matrix method
introduced in \cite{Borasoy:2006qn}.

Overall we find that the Gaussian-smeared actions LO$_{2}$ and NLO$_{2}$ are
more accurate than the standard actions LO$_{1}$ and NLO$_{1}$. \ This can be
seen most easily from the size of the required NLO corrections in Table
\ref{fitvalues}. \ For $P$-wave interactions, however, we see from comparing
Fig. \ref{pwave_b0} and Fig. \ref{pwave_b6} that the standard actions LO$_{1}$
and NLO$_{1}$ are more accurate. \ In future studies it would be useful to try
to find an improved LO action with accurate $S$-wave phase shifts and smaller
$P$-wave attraction without inducing sign problems in Monte Carlo simulations.

The results presented here can be extended to higher orders in chiral
effective field theory. \ If we continue with the low cutoff momentum
$\Lambda\approx2.3m_{\pi}$, then the NNLO two-pion exchange potential can be
expanded in powers of $q^{2}/(4m_{\pi}^{2})$ in the same manner as the NLO
two-pion exchange potential. \ This expansion yields local operators which are
either renormalizations of local operators with zero or two derivatives, local
operators with four derivatives, or local operators with more than four
derivatives. \ The operators with four derivatives should be treated in the
same manner as local four-derivative operators appearing at N$^{3}$LO in the
usual chiral power counting.

In our analysis we have ignored small effects that appear in extraneous
channels due to broken rotational invariance on the lattice. \ For example the
mixed $^{3}S_{1}$-$^{3}D_{1}$ spherical waves each have some small admixture
of an unphysical $^{3}D_{3}$ component. \ In our analysis of $\varepsilon_{1}$
we measured the ratio of $^{3}S_{1}$ and $^{3}D_{1}$ components and neglected
this small $^{3}D_{3}$ component. \ In the future we might prefer a more
ambitious approach which explicitly removes the $^{3}D_{3}$ component. \ This
requires including local interactions which are invariant under the cubic
rotational group SO$(3,\mathbb{Z})$ but not invariant under the full SO$(3)$
symmetry. \ For example in addition to the SO$(3)$-invariant interaction%
\begin{equation}
\frac{1}{2}:\sum\limits_{\vec{n}}\sum\limits_{S}\Delta_{S}\rho_{S}%
^{a^{\dagger},a}(\vec{n})\sum\limits_{S^{\prime}}\Delta_{S^{\prime}}%
\rho_{S^{\prime}}^{a^{\dagger},a}(\vec{n}):,
\end{equation}
we should also include a small contribution from the SO$(3,\mathbb{Z}%
)$-invariant but SO$(3)$-noninvariant operator%
\begin{equation}
\frac{1}{2}:\sum\limits_{\vec{n}}\sum\limits_{S}\Delta_{S}\rho_{S}%
^{a^{\dagger},a}(\vec{n})\Delta_{S}\rho_{S}^{a^{\dagger},a}(\vec{n}):.
\end{equation}

\section*{Acknowledgements}

Partial financial support from the Deutsche Forschungsgemeinschaft (SFB/TR
16), Helmholtz Association (contract number VH-NG-222 and VH-VI-231), and U.S.
Department of Energy (DE-FG02-03ER41260) are gratefully acknowledged. \ This
research is part of the EU Integrated Infrastructure Initiative in Hadron
Physics under contract number RII3-CT-2004-506078. \ The computational
resources for this project were provided by the John von Neumann Institute for
Computing at the Forschungszentrum J\"{u}lich.

\appendix

\section{Continuum limit of the NLO interactions}

\subsection{One-nucleon matrix elements}

The matrix elements of the local densities for single nucleon states are%
\begin{equation}
\left\langle \vec{p},i,j\right\vert \rho^{a^{\dagger},a}(\vec{n})\left\vert
\vec{p}\,^{\prime},i^{\prime},j^{\prime}\right\rangle =\delta_{ii^{\prime}%
}\delta_{jj^{\prime}}e^{i\left(  \vec{p}\,^{\prime}-\vec{p}\right)  \cdot
\vec{n}},
\end{equation}%
\begin{equation}
\left\langle \vec{p},i,j\right\vert \rho_{I}^{a^{\dagger},a}(\vec
{n})\left\vert \vec{p}\,^{\prime},i^{\prime},j^{\prime}\right\rangle
=\delta_{ii^{\prime}}\left[  \tau_{I}\right]  _{jj^{\prime}}e^{i\left(
\vec{p}\,^{\prime}-\vec{p}\right)  \cdot\vec{n}},
\end{equation}%
\begin{equation}
\left\langle \vec{p},i,j\right\vert \rho_{S}^{a^{\dagger},a}(\vec
{n})\left\vert \vec{p}\,^{\prime},i^{\prime},j^{\prime}\right\rangle =\left[
\sigma_{S}\right]  _{ii^{\prime}}\delta_{jj^{\prime}}e^{i\left(  \vec
{p}\,^{\prime}-\vec{p}\right)  \cdot\vec{n}},
\end{equation}%
\begin{equation}
\left\langle \vec{p},i,j\right\vert \rho_{S,I}^{a^{\dagger},a}(\vec
{n})\left\vert \vec{p}\,^{\prime},i^{\prime},j^{\prime}\right\rangle =\left[
\sigma_{S}\right]  _{ii^{\prime}}\left[  \tau_{I}\right]  _{jj^{\prime}%
}e^{i\left(  \vec{p}\,^{\prime}-\vec{p}\right)  \cdot\vec{n}}.
\end{equation}
The matrix elements for derivatives of these densities are found by taking
$f(\vec{n})=e^{i\left(  \vec{p}\,^{\prime}-\vec{p}\right)  \cdot\vec{n}}$ and
computing%
\begin{equation}
\Delta_{l}f(\vec{n})=e^{i\left(  \vec{p}\,^{\prime}-\vec{p}\right)  \cdot
\vec{n}}\times\frac{1}{4}\sum_{\substack{\nu_{1},\nu_{2},\nu_{3}%
=0,1}}(-1)^{\nu_{l}+1}e^{i\left(  \vec{p}\,^{\prime}-\vec{p}\right)  \cdot
\vec{\nu}},
\end{equation}%
\begin{equation}
\triangledown_{l}^{2}f(\vec{n})=e^{i\left(  \vec{p}\,^{\prime}-\vec{p}\right)
\cdot\vec{n}}\times\left[  e^{i\left(  \vec{p}\,^{\prime}-\vec{p}\right)
\cdot\hat{l}}+e^{-i\left(  \vec{p}\,^{\prime}-\vec{p}\right)  \cdot\hat{l}%
}-2\right]  .
\end{equation}
In the continuum limit $\left\vert \vec{p}\,^{\prime}\right\vert $ and
$\left\vert \vec{p}\right\vert $ in units of inverse lattice spacing are small
and so%
\begin{equation}
\frac{1}{4}\sum_{\substack{\nu_{1},\nu_{2},\nu_{3}=0,1}}(-1)^{\nu_{l}%
+1}e^{i\left(  \vec{p}\,^{\prime}-\vec{p}\right)  \cdot\vec{\nu}}\rightarrow
i\left(  p\,_{l}^{\prime}-p_{l}\right)  ,
\end{equation}%
\begin{equation}
e^{i\left(  \vec{p}\,^{\prime}-\vec{p}\right)  \cdot\hat{l}}+e^{-i\left(
\vec{p}\,^{\prime}-\vec{p}\right)  \cdot\hat{l}}-2\rightarrow-\left(
p\,_{l}^{\prime}-p_{l}\right)  ^{2}.
\end{equation}

The matrix element for the SU(4)-invariant current density is%
\begin{equation}
\left\langle \vec{p},i,j\right\vert \Pi_{l}^{a^{\dagger},a}(\vec{n})\left\vert
\vec{p}\,^{\prime},i^{\prime},j^{\prime}\right\rangle =\delta_{ii^{\prime}%
}\delta_{jj^{\prime}}e^{i\left(  \vec{p}\,^{\prime}-\vec{p}\right)  \cdot
\vec{n}}\times\frac{1}{4}\sum_{\substack{\nu_{1},\nu_{2},\nu_{3}%
=0,1}}(-1)^{\nu_{l}+1}e^{i\vec{p}\,^{\prime}\cdot\vec{\nu}}e^{-i\vec{p}%
\cdot\vec{\nu}(-l)}.
\end{equation}
Similarly the matrix element for the spin current density is%
\begin{equation}
\left\langle \vec{p},i,j\right\vert \Pi_{l,S}^{a^{\dagger},a}(\vec
{n})\left\vert \vec{p}\,^{\prime},i^{\prime},j^{\prime}\right\rangle =\left[
\sigma_{S}\right]  _{ii^{\prime}}\delta_{jj^{\prime}}e^{i\left(  \vec
{p}\,^{\prime}-\vec{p}\right)  \cdot\vec{n}}\times\frac{1}{4}\sum
_{\substack{\nu_{1},\nu_{2},\nu_{3}=0,1}}(-1)^{\nu_{l}+1}e^{i\vec{p}%
\,^{\prime}\cdot\vec{\nu}}e^{-i\vec{p}\cdot\vec{\nu}(-l)}.
\end{equation}
The other current densities are not needed for the NLO interactions. \ In the
continuum limit%
\begin{equation}
\frac{1}{4}\sum_{\substack{\nu_{1},\nu_{2},\nu_{3}=0,1}}(-1)^{\nu_{l}%
+1}e^{i\vec{p}\,^{\prime}\cdot\vec{\nu}}e^{-i\vec{p}\cdot\vec{\nu}%
(-l)}\rightarrow i\left(  p\,_{l}^{\prime}+p_{l}\right)  .
\end{equation}

\subsection{Two-nucleon matrix elements}

The tree-level amplitude for two-nucleon scattering consists of contributions
from direct and exchange diagrams. \ However for bookkeeping purposes we label
the contact interactions according to the tree-level amplitude for scattering
of two distinguishable nucleons. \ We imagine that one nucleon is of type $A$
and the other nucleon is of type $B$. \ The interactions include densities and
current densities for both $A$ and $B$. \ For example the SU(4)-invariant
density and current density become%
\begin{equation}
\rho^{a^{\dagger},a}(\vec{n})\rightarrow\rho^{a_{A}^{\dagger},a_{A}}(\vec
{n})+\rho^{a_{B}^{\dagger},a_{B}}(\vec{n}),
\end{equation}%
\begin{equation}
\Pi_{l}^{a^{\dagger},a}(\vec{n})\rightarrow\Pi_{l}^{a_{A}^{\dagger},a_{A}%
}(\vec{n})+\Pi_{l}^{a_{B}^{\dagger},a_{B}}(\vec{n}).
\end{equation}

Let the incoming momenta be $\vec{p}_{i}^{\,A}$ and $\vec{p}_{i}^{\,B}$ and
the outgoing momenta be $\vec{p}_{f}^{\,A}$ and $\vec{p}_{f}^{\,B}$. \ The
$t$-channel momentum transfer is%
\begin{equation}
\vec{q}=\vec{p}_{i}^{\,A}-\vec{p}_{f}^{\,A}=-\vec{p}_{i}^{\,B}+\vec{p}%
_{f}^{\,B},
\end{equation}
and the $u$-channel exchanged momentum transfer is%
\begin{equation}
\vec{k}=\vec{p}_{i}^{\,A}-\vec{p}_{f}^{\,B}=-\vec{p}_{i}^{\,B}+\vec{p}%
_{f}^{\,A}.
\end{equation}
For these incoming and outgoing states the amplitudes for $\Delta V^{(0)}$ are%
\begin{equation}
\mathcal{A}\left(  \Delta V\right)  =\Delta C,
\end{equation}%
\begin{equation}
\mathcal{A}\left(  \Delta V_{I^{2}}\right)  =\Delta C_{I^{2}}\sum_{I}\tau
_{I}^{A}\tau_{I}^{B}.
\end{equation}
In the continuum limit the amplitudes for $V^{(2)}$ are%
\begin{equation}
\mathcal{A}\left(  V_{q^{2}}\right)  \rightarrow C_{q^{2}}q^{2},
\end{equation}%
\begin{equation}
\mathcal{A}\left(  V_{I^{2},q^{2}}\right)  \rightarrow C_{I^{2},q^{2}}%
q^{2}\sum_{I}\tau_{I}^{A}\tau_{I}^{B},
\end{equation}%
\begin{equation}
\mathcal{A}\left(  V_{S^{2},q^{2}}\right)  \rightarrow C_{S^{2},q^{2}}%
q^{2}\sum_{S}\sigma_{S}^{A}\sigma_{S}^{B},
\end{equation}%
\begin{equation}
\mathcal{A}\left(  V_{S^{2},I^{2},q^{2}}\right)  \rightarrow C_{S^{2}%
,I^{2},q^{2}}q^{2}\sum_{S}\sigma_{S}^{A}\sigma_{S}^{B}\sum_{I}\tau_{I}^{A}%
\tau_{I}^{B},
\end{equation}%
\begin{equation}
\mathcal{A}\left(  V_{(q\cdot S)^{2}}\right)  \rightarrow C_{(q\cdot S)^{2}%
}\sum_{S}q_{S}\sigma_{S}^{A}\sum_{S^{\prime}}q_{S^{\prime}}\sigma_{S^{\prime}%
}^{B},
\end{equation}%
\begin{equation}
\mathcal{A}\left(  V_{I^{2},(q\cdot S)^{2}}\right)  \rightarrow C_{I^{2}%
,(q\cdot S)^{2}}\sum_{I}\tau_{I}^{A}\tau_{I}^{B}\sum_{S}q_{S}\sigma_{S}%
^{A}\sum_{S^{\prime}}q_{S^{\prime}}\sigma_{S^{\prime}}^{B},
\end{equation}%
\begin{equation}
\mathcal{A}\left(  V_{(iq\times S)\cdot k}\right)  \rightarrow iC_{(iq\times
S)\cdot k}\sum_{l,S,l^{\prime}}\varepsilon_{lSl^{\prime}}q_{l}\left(
\sigma^{A}+\sigma^{B}\right)  _{S}k_{l^{\prime}}.
\end{equation}
Most of these are straightforward, however the spin-orbit amplitude
$\mathcal{A}\left(  V_{(iq\times S)\cdot k}\right)  $ requires some
derivation. \ In terms of the incoming and outgoing momenta,%
\begin{align}
&  \mathcal{A}\left(  V_{(iq\times S)\cdot k}\right) \nonumber\\
&  \rightarrow\frac{i}{2}C_{(iq\times S)\cdot k}\sum\limits_{l,S,l^{\prime}%
}\varepsilon_{lSl^{\prime}}\left[  \left(  p_{i}^{A}+p_{f}^{A}\right)
_{l}\sigma_{S}^{B}\left(  p_{i}^{B}-p_{f}^{B}\right)  _{l^{\prime}}+\left(
p_{i}^{B}+p_{f}^{B}\right)  _{l}\sigma_{S}^{A}\left(  p_{i}^{A}-p_{f}%
^{A}\right)  _{l^{\prime}}\right] \nonumber\\
&  +\frac{i}{2}C_{(iq\times S)\cdot k}\sum\limits_{l,S,l^{\prime}}%
\varepsilon_{lSl^{\prime}}\left[  \left(  p_{i}^{A}+p_{f}^{A}\right)
_{l}\sigma_{S}^{A}\left(  p_{i}^{B}-p_{f}^{B}\right)  _{l^{\prime}}+\left(
p_{i}^{B}+p_{f}^{B}\right)  _{l}\sigma_{S}^{B}\left(  p_{i}^{A}-p_{f}%
^{A}\right)  _{l^{\prime}}\right]  .
\end{align}
We note that%
\begin{align}
\sum_{l,S,l^{\prime}}\varepsilon_{lSl^{\prime}}\left(  p_{i}^{A}+p_{f}%
^{A}\right)  _{l}\sigma_{S}^{B}\left(  p_{i}^{B}-p_{f}^{B}\right)
_{l^{\prime}}  &  =-\sum_{l,S,l^{\prime}}\varepsilon_{lSl^{\prime}}\left[
\left(  p_{i}^{A}+p_{f}^{A}\right)  _{l}\sigma_{S}^{B}\left(  p_{i}^{A}%
-p_{f}^{A}\right)  _{l^{\prime}}\right] \nonumber\\
&  =-\sum_{l,S,l^{\prime}}\varepsilon_{lSl^{\prime}}\left[  \left(  2p_{i}%
^{A}\right)  _{l}\sigma_{S}^{B}\left(  p_{i}^{A}-p_{f}^{A}\right)
_{l^{\prime}}\right]  . \label{one}%
\end{align}
Similarly%
\begin{equation}
\sum_{l,S,l^{\prime}}\varepsilon_{lSl^{\prime}}\left(  p_{i}^{B}+p_{f}%
^{B}\right)  _{l}\sigma_{S}^{A}\left(  p_{i}^{A}-p_{f}^{A}\right)
_{l^{\prime}}=-\sum_{l,S,l^{\prime}}\varepsilon_{lSl^{\prime}}\left[  \left(
2p_{f}^{B}\right)  _{l}\sigma_{S}^{A}\left(  p_{i}^{B}-p_{f}^{B}\right)
_{l^{\prime}}\right]  , \label{two}%
\end{equation}%
\begin{equation}
\sum_{l,S,l^{\prime}}\varepsilon_{lSl^{\prime}}\left(  p_{i}^{A}+p_{f}%
^{A}\right)  _{l}\sigma_{S}^{A}\left(  p_{i}^{B}-p_{f}^{B}\right)
_{l^{\prime}}=-\sum_{l,S,l^{\prime}}\varepsilon_{lSl^{\prime}}\left[  \left(
2p_{i}^{A}\right)  _{l}\sigma_{S}^{A}\left(  p_{i}^{A}-p_{f}^{A}\right)
_{l^{\prime}}\right]  , \label{three}%
\end{equation}%
\begin{equation}
\sum_{l,S,l^{\prime}}\varepsilon_{lSl^{\prime}}\left(  p_{i}^{B}+p_{f}%
^{B}\right)  _{l}\sigma_{S}^{B}\left(  p_{i}^{A}-p_{f}^{A}\right)
_{l^{\prime}}=-\sum_{l,S,l^{\prime}}\varepsilon_{lSl^{\prime}}\left[  \left(
2p_{f}^{B}\right)  _{l}\sigma_{S}^{B}\left(  p_{i}^{B}-p_{f}^{B}\right)
_{l^{\prime}}\right]  . \label{four}%
\end{equation}
The sum of the four terms in Eq.~(\ref{one})-(\ref{four}) gives%
\begin{equation}
\sum_{l,S,l^{\prime}}\varepsilon_{lSl^{\prime}}\left[  -\left(  2p_{i}%
^{A}\right)  _{l}\sigma_{S}^{B}q_{l^{\prime}}+\left(  2p_{f}^{B}\right)
_{l}\sigma_{S}^{A}q_{l^{\prime}}-\left(  2p_{i}^{A}\right)  _{l}\sigma_{S}%
^{A}q_{l^{\prime}}+\left(  2p_{f}^{B}\right)  _{l}\sigma_{S}^{B}q_{l^{\prime}%
}\right]  ,
\end{equation}
which equals%
\begin{equation}
\sum_{l,S,l^{\prime}}\varepsilon_{lSl^{\prime}}\left[  -2k_{l}\sigma_{S}%
^{B}q_{l^{\prime}}-2k_{l}\sigma_{S}^{A}q_{l^{\prime}}\right]  .
\end{equation}
We therefore find%
\begin{align}
\mathcal{A}\left(  V_{(iq\times S)\cdot k}\right)   &  =-iC_{(iq\times S)\cdot
k}\sum_{l,S,l^{\prime}}\varepsilon_{lSl^{\prime}}\left[  k_{l}\sigma_{S}%
^{B}q_{l^{\prime}}+k_{l}\sigma_{S}^{A}q_{l^{\prime}}\right] \nonumber\\
&  =iC_{(iq\times S)\cdot k}\sum_{l,S,l^{\prime}}\varepsilon_{lSl^{\prime}%
}q_{l}\left(  \sigma^{A}+\sigma^{B}\right)  _{S}k_{l^{\prime}}.
\end{align}

\bibliographystyle{apsrev}
\bibliography{NuclearMatter}

\end{document}